\documentclass[aps,pre,twocolumn,showpacs,floatfix,superscriptaddress]{revtex4}

\usepackage{epsfig}
\usepackage{graphicx}

\begin{document}

\title{Networks of Recurrent Events, a Theory of Records,
and an Application to Finding Causal Signatures in Seismicity}

\author{J\"orn Davidsen}
\email{davidsen@phas.ucalgary.ca}
\affiliation{Complexity Science Group, Department of
Physics and Astronomy, University of Calgary, Calgary, Alberta,
Canada  T2N 1N4}

\author{Peter Grassberger}
\affiliation{Complexity Science Group, Department of
Physics and Astronomy, University of Calgary, Calgary, Alberta,
Canada  T2N 1N4}

\affiliation{Institute for Biocomplexity and Informatics,
University of Calgary, Calgary, Alberta, Canada}

\author{Maya Paczuski}
\affiliation{Complexity Science Group, Department of
Physics and Astronomy, University of Calgary, Calgary, Alberta,
Canada  T2N 1N4}

\date{\today}

\begin{abstract}
We propose a method to search for signs of causal structure in
spatiotemporal data making minimal \emph{a priori} assumptions
about the underlying dynamics. To this end, we generalize the
elementary concept of recurrence for a point process in time to
recurrent events in space and time. An event is defined to be a
recurrence of any previous event if it is closer to it in space
than all the intervening events. As such, each sequence of
recurrences for a given event is a record breaking process. This
definition provides a strictly data driven technique to search for
structure. Defining events to be nodes, and linking each event to
its recurrences, generates a network of recurrent events.
Significant deviations in statistical properties of that network compared to
networks arising from (acausal) random processes allows one to
infer attributes of the causal dynamics that generate observable
correlations in the patterns. We derive analytically a number of
properties for the network of recurrent events composed by a
random process in space and time. We extend the theory of records
to treat not only the variable where records happen, but also time
as continuous. In this way, we construct a fully symmetric theory
of records leading to a number of new results. Those analytic
results are compared in detail to the properties of a network
synthesized from time series of epicenter locations for
earthquakes in Southern California. Significant disparities from
the ensemble of acausal networks that can be plausibly attributed
to the causal structure of seismicity are: (1) Invariance of
network statistics with the time span of the events considered,
(2) Appearance of a fundamental length scale for recurrences,
independent of the time span of the catalog, which is consistent
with observations of the ``rupture length'', (3) Hierarchy in the
distances and times of subsequent recurrences. As expected, almost
all of the statistical properties of a network constructed from a
surrogate in which the original magnitudes and locations of
earthquake epicenters are randomly ``shuffled'' are completely
consistent with predictions from the acausal null model.

\end{abstract}

\pacs{02.50.-r,05.65.+b,91.30.Dk,05.45.Tp}

\maketitle

\section{Introduction}

Many striking features of physical, biological or social processes
can be portrayed as patterns or clusters of localized events.
These can be flips of magnetic domains in a ferromagnet leading to
Barkhausen noise~\cite{cote91,sethna01}, traffic
jams~\cite{nagel95}, booms and busts of markets and
economies~\cite{bak,farmer05}, forest fires~\cite{bak90}, the spread of
infections~\cite{turcotte99} and global pandemics, extinctions of
species~\cite{bak,bak93,crutchfield,drossel01}, neural
spikes~\cite{softky92}, solar flares~\cite{hughes03,paczuski04},
or earthquakes \cite{turcotte,stein,baiesi05} -- to name a few. A
generic attribute in all these cases is that one event can trigger
or somehow induce another one to occur -- or possibly numerous
further events. Sometimes, as in the prototype sandpile
model~\cite{bak87}, an accounting of causes and their effects
leads to an interpretation in terms of avalanches - where causal
connections between clustered events (``topplings'') are
explicitly rationalized by the microscopic state and rules of the
dynamical system. More often than not, though, the network of
causal connections cannot be resolved from the data at hand and
remains ambiguous. Thus, one is often confronted with inferring a
plausible causal structure from clusters of localized events
without a detailed or ``fundamental'' knowledge of the true
microscopic dynamics. This remains a stubbornly impenetrable
problem despite some progress in special cases (see e.g.
Ref.~\cite{marzocchi04} and references therein).

We aim to establish a general procedure of plausible inference
based on sequences of data in space and time, or more generally
for any temporal sequence of data. The essential idea for the
method of analysis discussed here is that of a \emph{recurrence}.
Our definition of recurrences is a generalization of ``returns''
for a point process to higher dimensional data structures that
evolve in time. Loosely spoken, a recurrence involves a pair of
events which are sufficiently close to each other
to suggest a causal connection.

\subsection{An example of contextual dependence}

For illustration consider the two events: $(A)$ First, Alice drops
a banana, and ($B$) then Bob falls down. If $A$ and $B$ are
sufficiently close in space and time then one can reasonably infer
that it is likely that Bob slipped on the banana and fell down
(``$A$ caused $B$''), but should these events be sufficiently
separated then $A$ is less likely to have contributed to $B$'s
occurrence. For instance, Bob could have been distracted by the
banana, or fell for another reason related to $A$ without actually
slipping directly on the banana -- so the two events may still be
connected without $A$ being exclusively the cause of $B$. This
secondary effect is also less likely if sufficient time has past
between the two events. Eventually Alice or another party may pick
up the banana or Bob's fall may have happened so far away that it
would be unlikely for him to have slipped on it.

As this example shows, it is not always clear what we should
mean by 'sufficiently close' to infer a causal connection. One option
might be to call a localized event $B$ a recurrence of an earlier
event $A$, if its spatial distance is less than some
chosen length $l$ \cite{note1}.
In addition to introducing a length scale, this choice fails to
admit that the plausibility of causal connections typically
becomes weaker with time -- as the example above makes plain. In
addition, the likelihood that the later event ($B$) may be
triggered by a third intervening event increases with time as
well. These considerations might suggest that $l$ should shrink
with time. On the other hand, the fact that influences usually
spread either diffusively or with finite speed could suggest the
opposite - that $l$ increases with time. Spreading of influence is
hypothesized, for instance, in theories of ``aftershock zone
diffusion''~(see Ref.~\cite{helmstetter03b} and references
therein). Other, more complicated scenarios are also conceivable.

This discussion is meant to clarify that without sufficiently
accurate \emph{a priori} knowledge of the underlying microscopic
dynamics any definition of closeness based on predefined scales is
arbitrary and might significantly alter the inferred causal
structure. To avoid this problem, or more generally to minimize
the influence of the observer, we take the view that, to begin
with, a suitable definition of closeness ought to be purely
contextual, and depend only on the actual history of events.
Taking this as our starting point -- that we know the observed
history of events but do not know the underlying dynamics -- we
propose a contextual method to establish recurrences that uses
'zero knowledge' of the underlying physical processes. As a
result, our definition is generic and can apply to a wide variety
of situations. This approach serves as a starting point to analyze
data for systems where the underlying dynamics is obscure,
mysterious or even misconceived. It comprises a fundamental
extension of the concept of recurrences for a point process to
recurrent events in space and time that allows the inference of
causal relations from available or possible observations.

\subsection{Contextual relationships represented by a network}

In the approach described here, the inferred relationship between
each pair of events is based on the closeness of the pair relative
to all the other events that have occurred in the data set. An
event $B$ is designated to be a recurrence of a previous one $A$
if it is closer to $A$ --- compared to any other event occurring
in the time interval between $A$ and $B$. By this construction,
each recurrence is a new ``record'' in the sequence of distances
that subsequent events have from $A$. In other words, each
recurrence is a \emph{record breaking
event}~\cite{glick78,nevzorov87,nevzorov}.

This method of inferring relationships between pairs of events is
naturally expressed as a network of connected events where each
event is a node in the graph, and each recurrent pair is linked
with a time directed edge. Significant deviations in the
statistics of the resulting network from that for a random process
(which lacks any causal relations between events) highlights
relevant parts of the causal dynamical process(es) generating the
patterns. In principle, the events themselves do not have to take
place in real physical space, but can occur in any space as long
as it is equipped with a metric that defines distances. As a
starting point, here we only discuss spatiotemporal point
processes and take as our test bed a well-characterized, extensive
and comparatively accurate catalog~\cite{catalog} of earthquake
epicenters for Southern California.

\subsection{Outline}

Section II explains our method for constructing networks of
recurrent events and the relation to record breaking statistics.
In Section III, the null hypothesis of independent, random events
is introduced and a number of analytic results are obtained for
it. We extend the mathematical theory of record breaking
statistics to the case where both space (or the variable which
fluctuates and in which records take place) and time (or the
ordering of events) are treated on the same footing. Treating both
space and time as continuous symmetrizes the theory -- making it
more concise. These results allow us to discover statistical
features in the actual network of recurrences that are unlikely in
acausal random processes and, hence, plausibly due to causal
structures in the underlying dynamics.

Section IV describes the application to seismicity. The network
analysis reveals new statistical features of seismicity --- with
robust scaling laws that are invariant over a range of different
time scales. This apparent invariance with respect to the time
span is diametrically opposed to the behavior for a random
process, where all statistical distributions depend explicitly on
the time span over which events are recorded. The rupture length
and its scaling with magnitude (while being invariant with respect
to the time span of the history) emerges from the data analysis
without being predefined by the measurement process. It is a
generic measure for distance between recurrent events. These
results indicate that our method is, indeed, tending to identify
causally related events rather than acausal pairs. Further, the
relative separations for subsequent recurrences in space (or time)
form a hierarchy with unexpected properties. All of these
properties disappear when a history constructed by ``shuffling''
the original earthquake catalog is analyzed using the same method.
In that case, almost all results agree with predictions of the
acausal null model. On the basis of these results, we argue that
the particular features where we observe strong deviations between
the actual history and the acausal null model can be attributed to
causal structures in the dynamics of seismicity. We end with a
summary and outlook for future works and applications.

\section{Synthesizing the Network of Recurrences\label{sec:synth}}

Consider a series of events $a_i$, with $i=1 \ldots N$, that are
ordered in time such that event $a_i$ precedes event $a_j$ if
$i<j$. The events $a_i$ are in the following identified with their
spatiotemporal position. We assume that a metric is defined in
space, and we denote by $d_{ij}$ the spatial distance between
events $a_i$ and $a_j$. Simple examples are spatiotemporal point
processes taking place in 3-dimensional Euclidean space or on the
surface of a sphere. The only property of the metric relevant to
this discussion is that (spatial) distances between all pairs of
events can be ordered, e.g. from smallest to largest, and the
ordering relation is transitive. The same is of course true for
time distances.

\begin{figure}[htbp]
\includegraphics*[width=0.8\columnwidth]{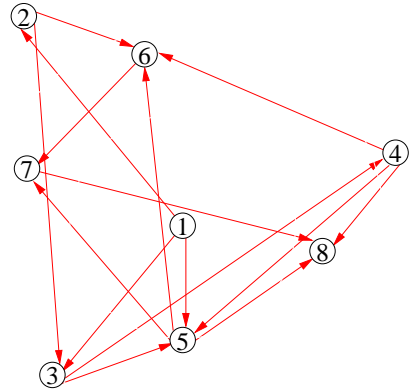}
\caption{\label{cascade} Eight events in 2D space labelled
according to their order of occurrence in time. The network of
recurrences is indicated by arrows as described in the text.}
\end{figure}

The network of recurrent events is defined as follows (see
Fig.~\ref{cascade}): All events $a_i$ are represented as nodes and
two nodes $i$ and $j$ with $i<j$ are connected by a directed link
or edge $e_{ij}$ if event $a_j$ is a recurrence of $a_i$. This
occurs if and only if $d_{ij} < d_{ik}$ for all $k$ with $i<k<j$.
Thus a recurrence is a new record with respect to distance. Note
that $e_{ij}$ and $e_{ji}$ cannot both exist since the
directionality of links is determined by the time ordering. Hence,
if $i<j$ only $e_{ij}$ can exist. To summarize: the definition of
recurrence implies that, for all $2<j<n$, event $a_{j}$ is
automatically a recurrence of event $a_{j-1}$ and, thus, all links
$e_{(j-1)j}$ exist. Event $a_{j}$ is also a recurrence of any
previous event $a_i$ if it is closer to $a_i$ than every other
event $a_k$ that occurred in between the two, i.e., for all $a_k$
with $i<k<j$.

As long as only one event occurs at a time, the directed network
consists of a single cluster in which each node is linked to at
least one other node. Each node $i$ has an in-degree $k^{in}_i$,
which is the number of links pointing to it from events in its
past, as well as an out-degree $k^{out}_i$, which is the number of
links emanating from $i$ -- corresponding to the number of records
of event $i$. The collection of in-nodes $I_i=\{j \mid e_{ji} \;
\mathrm{exists}\}$ are hypothesized to reflect the potential
cause(s) of event $a_i$ while the set of out-nodes $O_i=\{j \mid
e_{ij} \; \mathrm{exists}\}$ are hypothesized to contain the
effect(s) of $a_i$. Although it is natural to contemplate
associating a weight factor to each link, this requires further
assumptions. Here we do not deal with this issue and consider all
links to have the same weight. This is in our view a ``zeroth
order'' assignment of causes and their effects based purely on the
history of events and their relationships to each other in space
and time. Note that a single event can have
many causes corresponding to all of its incoming links, so the
network aspect of causal relations is not lost in this limit.
Weighted networks of seismic events were constructed using a
different methodology in Refs.~\cite{baiesi04,baiesi05,baiesi06}.

While this network construction, based on record breaking events,
is directly applicable to fixed collections of events, it can also
be applied when the number of events $N$ increases over time. The
result of adding a new event $a_{N+1}$ is to increase the number
of links by at least one, namely $e_{N(N+1)}$, without altering
any pre-existing links. Hence, the property of being a recurrence
is \emph{preserved} in all cases under addition of new nodes in
time. Also the collections of in-nodes for all pre-existing nodes
remains unchanged. Yet, the out-degree of any node $i$ with
$i<N+1$ can increase by one, namely if $a_{N+1}$ is a recurrence
of $a_i$. So the networks are, in this sense, dynamically stable
growing networks~\cite{albert02,newman03}.

Some tools and measures already exist to quantify statistical
topological features of networks, and to reveal the organization
of the dynamical process(es) giving rise to the events in terms of
network statistics \cite{albert02,newman03}. The dependence of the
network statistics can be examined by varying the time span of the
history synthesized into a network, space window over which the
history is observed, and/or selection criteria for what is defined
as an event (in the seismic application discussed later, this
could e.g. be the range of earthquake magnitudes). Our approach
opens up a new view of dynamical organization of spatiotemporal
activity in terms of the (static) topology of complex networks --
as was also discussed
in~\cite{baiesi04,baiesi05,baiesi06,davidsen05pm,shreim06}. We
also believe it possible that new developments in network theory
may turn out to be even more powerful in analyzing dynamical
systems. For the work described here, standard methods of network
analysis are already sufficient to plausibly infer certain causal
relations in seismic behavior solely from the catalog of
earthquake magnitudes, epicenter locations and times.

\section{The acausal null model and a theory of records\label{sec:null}}

\subsection{General remarks}

In order to be able to associate causal characteristics of the
dynamics to the network of recurrences, we mathematically
establish statistical properties of a null model, where the events
in space and time are random, uncorrelated and causally unrelated.
Then any statistically significant deviation of the observed
network from this null hypothesis can be attributed to
correlations among events and to causal structure in the
underlying dynamics giving rise to the observed history. The
conclusions about the relation to causality are robust as long as
the relevant properties of any acausal null model are well
represented by those we study.

In the following we shall discuss several variants of the null
model. In all of them, both space and time are continuous. To the
best of our knowledge, the theory of records has up to now been
developed only for discrete time and continuous space
\cite{glick78,nevzorov87,nevzorov}. As we shall see, when both
variables are continuous the core of the theory becomes symmetric
under exchange of space and time, allowing for a more concise
formulation. This symmetry is obviously lost when making one of
the variables discrete.

Let us denote by $\rho({\bf x}_1,t_1; \ldots {\bf x}_n,t_n)$ the
joint probability density for having events at locations $({\bf
x}_i,t_i)$, $i=1,\ldots n$. Our basic assumptions are that:

(a) Events are independent and identically distributed (iid),
\begin{equation}
\rho_n({\bf x}_1,t_1; \ldots {\bf x}_n,t_n) = \prod_{i=1}^n
\rho_1({\bf x}_i,t_i).
\end{equation}

(b) The single-event distributions factorize,
\begin{equation}
\rho_1({\bf x},t) = \rho_x({\bf x})\rho_t(t).\label{factorize}
\end{equation}
In particular, when $\rho_t(t)=const$, Eq.~(\ref{factorize}) means
that we have a stationary system. Note that $\int d{\bf x} dt
\rho_1({\bf x},t) = N$, the total average number of events in the
history, as long as this number is finite.

Instead of event distributions themselves, we shall in the
following use the distributions of space-time distances relative
to some reference event or ``Event-0'' at $({\bf x}_0,t_0)$,
\begin{eqnarray}
\mu_n(l_1,t_1;\ldots l_n,t_n) & & = \prod_{i=1}^n
[\int d{\bf y}_i \delta(|{\bf x_0}-{\bf y}_i|-l_i)] \times\\
 & & \rho_n({\bf x}_0,t_0; {\bf y}_1,t_0+t_1; \ldots {\bf y}_n,t_0+t_n).
\nonumber
\end{eqnarray}
It is easily seen that these joint distributions also factorize
under the above assumptions as,
\begin{equation}
\mu_n(l_1,t_1;\ldots l_n,t_n) = \prod_{i=1}^n \mu_1(l_i,t_i)
\end{equation}
with
\begin{equation}
\mu_1(l,t) = \mu_l(l)\mu_t(t).
\end{equation}
The functions $\mu_l(l)$ and $\mu_t(t)$ might in general depend on
the reference point, ${\bf x_0}$. We will not indicate this
dependence explicitly, unless it is relevant for the calculation.

First, we consider the special case $\mu_l(l) = \mu_t(t) = 1$,
which holds if the system is stationary, 1-dimensional,
homogeneous, and has the suitable space-time density of events.
The next step is when either one of these functions or both are
equal to one up to finite cut-offs and zero beyond, i.e. $\mu_l(l)
= \Theta(\lambda-l)$ and/or $\mu_t(t) = \Theta(\sigma-t)$.
Physically, $\lambda$ is not only the maximal possible distance
between two events (due to finiteness of space), but it is also
the rate at which events occur per unit time, if $\mu_t(t)=1$.
Similarly, a finite value of $\sigma$ indicates not only that
events are observed in a finite time window, but also that the
average number of events per unit distance is finite.

\subsection{Canonical coordinates}

Fortunately, it is sufficient to discuss these simple cases,
because for any non-singular densities $\mu_l(l)$ and $\mu_t(t)$
the problem can be reduced to one of them by a change of
coordinates. Consider the two transformations
\begin{equation}
\xi = \int_0^l dl' \mu_l(l')\;,\qquad \tau = \int_0^t dt'
\mu_t(t')\;. \label{standard}
\end{equation}
Clearly, $\xi$ is a positive and monotonically increasing function
of $l$, while $\tau$ is a positive and monotonically increasing
function of $t$. Due to conservation of probability, both have
unit density
\begin{equation}
    \mu_\xi(\xi) = \Theta(\lambda-\xi) \;,\qquad \mu_\tau(\tau) =
    \Theta(\sigma-\tau),\label{equ:density}
\end{equation}
where we have denoted by $\lambda$ and $\sigma$ the integrals over
$\mu_l$ and $\mu_t$, respectively,
\begin{equation}
\lambda = \int_0^\infty dl'\mu_l(l')\;,\qquad \sigma =
\int_0^\infty dt' \mu_t(t').
\end{equation}
Thus, the distributions of events in $\xi$ and $\tau$ are cut-off
sharply at $\lambda$ and $\sigma$, respectively. Note that
$\lambda$ and $\sigma$ can be infinite.

Thus, for general space and time distributions, we can first do
all calculations in the ``canonical coordinates'' $\xi$ and $\tau$,
and then translate the results, using inverse transformations of
Eq.~(\ref{standard}), back to the original coordinates $l,t$.
Examples are given below. In the following we always assume that
$\xi$ and $\tau$ are defined by Eq.~(\ref{standard}) and, thus,
Eq.~(\ref{equ:density}) holds for all positive $\xi$ and $\tau$.

\begin{figure}[htbp]
\includegraphics*[width=0.7\columnwidth]{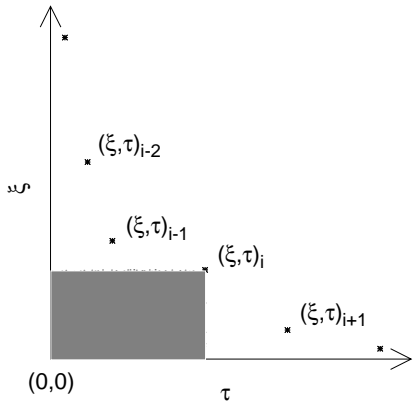}
\caption{\label{recurrence-chain} A typical chain of recurrences
in canonical coordinates. The reference event or Event-0 for all
these recurrences is at the origin $\tau=\xi=0$. The event at
$(\xi_i,\tau_i)$ is a recurrence of the event at (0,0) if and only
if no event is in the shaded region.}
\end{figure}

In canonical coordinates, a typical sequence of recurrences is
drawn schematically in Fig.~\ref{recurrence-chain}. For all
recurrences $i$, $\xi_{i+1} < \xi_i$ and $\tau_{i+1} > \tau_i$.
This is symmetric under the exchange $\xi \leftrightarrow
\tau,\;\lambda \leftrightarrow \sigma\;$, and $i \leftrightarrow
-i$. The probability that a given event $(\xi,\tau)$ is a
recurrence of Event-0 at (0,0) is equal to the chance that no
event occurred in the rectangular region $[0,\tau]\times [0,\xi]$,
which is equal to $\exp(-\xi\tau)$ due to the unit space-time
density of events in the $(\xi,\tau)$-plane. Hence, the joint
probability density function (PDF) of \emph{recurrences} is given
by the same exponential,
\begin{equation}
p(\xi,\tau) = e^{-\xi\tau},\label{2d-recur}
\end{equation} except for the possible cut-offs at $\lambda$ and/or
$\sigma$, beyond which the density of recurrences is zero; $p(\xi
> \lambda, \tau) = p(\xi, \tau > \sigma)= 0$.

\subsection{Infinite space and time domains}

For a detailed discussion of the spatial and temporal
distributions of recurrences we deal separately with the cases of
finite and infinite $\lambda$ and/or $\sigma$. We first consider
the case where neither $\mu_l(l)$ nor $\mu_t(t)$ is normalizable,
i.e. $\lambda=\sigma=\infty$. This, for example, describes the
case of stationary and homogeneous systems in infinite
$D$-dimensional Euclidean space, where $\mu_t(t)=const$ and
$\mu_l(l) \propto l^{D-1}$. But it holds also approximately for
fractal distributions in space (if we neglect effects of
lacunarity \cite{mandelbrot95}), with $D$ being the fractal
dimension. Notice that $\int_0^\infty dl l^{D-1}=\infty$ for all
values of $D$.

The spatial and temporal density distributions of recurrences in
canonical coordinates are obtained by integrating Eq.~(\ref{2d-recur}) to
obtain the marginals,
\begin{eqnarray}
 p_\xi(\xi) &=& \int_0^\infty d\tau\; p(\xi,\tau) = 1/\xi\;, \label{pxi} \\
 p_\tau(\tau) &=& \int_0^\infty d\xi\; p(\xi,\tau) = 1/\tau\;. \label{ptau}
\end{eqnarray}
Assuming that the system is translationally invariant in time and fractal in
space, i.e. $\mu_t(t) = b = const$ and $\mu_l(l) = aD l^{D-1}$,
we obtain for the densities in the original coordinates
\begin{eqnarray}
 p_l(l) & = & \mu_l(l) p_\xi(\xi(l)) = {aD l^{D-1} \over al^D} = D/l\;, \label{pl} \\
 p_t(t) & = & \mu_t(t) p_\tau(\tau(t)) = b [bt]^{-1} = 1/t\;. \label{pt}
\end{eqnarray}

Thus the recurrence density in time is independent of the event rate (per
unit space-time region). Similarly, for an event distribution with given
(fractal or Euclidean) non-trivial dimension, the recurrence density
depends on the dimension but not on the parameter $a$. Also, notice that
$p_t(t)$ is completely independent of the spatial event distribution
$\mu_l(l)$, and $p_l(l)$ is independent of $\mu_t(t)$.

For homogeneous and mono-fractal stationary spatial distributions
both $p_t$ and $p_l$ are independent of the reference point
defining the recurrences. This is no longer true for
multifractals, where $p_l(l)$ depends on the local (point-wise)
dimension at the event which defines the recurrences.

The analog of Eq.~(\ref{pt}) for discrete time is a classic result in
the theory of records \cite{glick78,nevzorov87,nevzorov}. In contrast,
Eq.~(\ref{pl}) was first reported in \cite{davidsen05pm}, as far as
we know.

\subsection{Finite space and infinite time --- and vice versa\label{sec:fiinfi}}

Let us assume that $\mu_t(t)$ is not normalizable but $\mu_l(l)$ is,
\begin{equation}
\lambda < \infty\;,\qquad \sigma = \infty\;.
\end{equation}

Now, of course, $p_\xi(\xi)=0$ for $\xi>\lambda$. For
$\xi<\lambda$, on the other hand, $p_\xi(\xi)$ is still given by
integrating $\exp(-\xi\tau)$ over all positive values of $\tau$ as
in Eq.~(\ref{pxi}), i.e.
\begin{equation}
p_\xi(\xi) = {1\over \xi} \Theta(\lambda-\xi) \;.
\end{equation}
In terms of the original coordinates, one finds
\begin{equation}
p_l(l) = \mu_l(l) p_\xi(\xi(l)) = {\mu_l(l)\over \int_0^l
dl'\;\mu_l(l')}.
\end{equation}

In contrast, $p_\tau(\tau)$ is obtained by integrating
Eq.~(\ref{2d-recur}) over the finite domain $0<\xi<\lambda$, which
gives
\begin{equation}
p_\tau(\tau) = {1\over \tau} (1-e^{-\lambda\tau}).\label{p_t}
\end{equation}
In the stationary case, when $t$ is just proportional to $\tau$,
the density of recurrences in $t$ is given by the same formula
with $\lambda$ replaced by the rate of events per unit $t$. The
additional term compared with Eqs.~(\ref{ptau}) and (\ref{pt})
reflects the probability that no recurrence occurs up to time
$\tau$ and $t$, respectively.

In the opposite case $(\lambda < \infty$, $\sigma=\infty)$ of
finite event rate per unit distance and infinite rate per unit
time (corresponding typically to infinite space and finite time,
with finite space time density of events), the situation is
completely symmetric. In that case $p_\tau(\tau)$ is cut-off
sharply at a finite value, while $p_\xi(\xi)$ is cut-off with an
exponential correction term as in Eq.~(\ref{p_t}).

\subsection{Finite space and finite time}

Now both $p_\xi(\xi)$ and $p_\tau(\tau)$ are obtained by
integrating Eq.~(\ref{2d-recur}) over finite domains,
\begin{eqnarray}
 p_\xi(\xi) & = & {1\over \xi} \Theta(\lambda-\xi) (1-e^{-\sigma\xi}),
 \label{pll} \\
 p_\tau(\tau) &=& {1\over \tau} \Theta(\sigma-\tau) (1-e^{-\lambda\tau}).
 \label{ptt}
\end{eqnarray}

Thus, $p_\xi(\xi)$ asymptotically approaches the constant $\sigma$
in the limit $\xi \to 0$ while for intermediate arguments we
recover the $1/\xi$ decay for infinite space and time domains
given in Eq.~(\ref{pxi}). For large arguments, the density sharply
drops to zero at $\xi=\lambda$. $p_\tau(\tau)$ asymptotically
approaches the constant $\lambda$ in the limit $\tau \to 0$ while
for intermediate arguments we recover the $1/\tau$ decay for
infinite space and time domains given in Eq.~(\ref{ptau}). For
large arguments, the density sharply drops to zero at
$\tau=\sigma$. The respective transition points $\xi^*$ and
$\tau^*$ between the constant behavior for small arguments and the
decaying behavior for intermediate arguments can be defined in the
standard way by requiring that the argument of the exponential
equals $-1$, i.e.,
\begin{eqnarray}
  \sigma \xi^* &\equiv& 1, \label{xi*}\\
  \lambda \tau^* &\equiv& 1. \label{tau*}
\end{eqnarray}

Specific realizations of such a process include stationary systems
observed over a finite time window, where events occur only in a
finite region of space --- or are only recorded when they fall
into that region. One example is $\mu_t(t) = b \Theta(T-t)$ and
$\mu_l = a D l^{D-1} \Theta(R-l)$ with positive constants $T$ and $R$.
In this case, Eq.~(\ref{pll}) translates into
\begin{equation}
 p_l(l) = \left\{\begin{array}{ll}
 a b T D l^{D-1} \quad & \mbox{ for } l \ll l^*(T)\mbox{, } l< R\;, \\
 D/l& \mbox{ for } l \gg l^*(T)\mbox{, } l< R\;, \\
0& \mbox{ for } l> R\;,
\end{array}\right. \label{model_p_l}
\end{equation}
and Eq.~(\ref{ptt}) translates into
\begin{equation}
 p_t(t) = \left\{\begin{array}{ll}
 a b R^D \quad & \mbox{ for } t \ll t^*(L)\mbox{, } t< T\;, \\
 1/t& \mbox{ for } t \gg t^*(L)\mbox{, } t< T\;, \\
0&\mbox{ for } t> T\;,
\end{array}\right. \label{model_p_t}
\end{equation}
with
\begin{equation}
l^*(T) \equiv (a b T)^{-1/D}\;,\label{l-cutoff-D}
\end{equation}
and
\begin{equation}
t^*(L) \equiv [a b R^D]^{-1}\;.\label{t-cutoff-D}
\end{equation}
Finally, let $\langle N\rangle =a b T R^D$ be the average
total number of observed events. Then the expressions for the
transition points are particularly simple
\begin{eqnarray}
  l^*(N) &=& L/\langle N\rangle^{1/D}, \\
  t^*(N) &=& T/\langle N\rangle. \label{t_N}
\end{eqnarray}

In this simple example and in the situations discussed in
subsection \ref{sec:fiinfi}, we have assumed that translational
invariance holds. However, this is generally not true. Specific
realizations of such processes include stationary systems observed
over a \emph{fixed} finite time window, where events occur only in
a \emph{fixed} finite region of space --- or are only recorded
when they fall into that region. Due to the lack of translational
invariance, the distributions of distances (spatial and temporal)
between events depend on the defining event. We discuss the
consequences of broken translational invariance now.

For concreteness and simplicity, let us assume a stationary
system where events occur uniformly on an interval $0 < x < L$
with periodic boundary conditions, with space-time density
$\alpha$. They are recorded only in the time window $0 < t < T$.
In general, the distributions of distances between events in a
bounded space-time region depend on the reference point
$(x_0,t_0)$, but in the present case this simplifies due to the
periodic boundary condition: The recurrence distributions depend
on $t_0$, but not on $x_0$. More precisely,
\begin{eqnarray}
 \mu_l(l;x_0,t_0) & = & 2\alpha \Theta(L/2-l), \\
 \mu_t(t;x_0,t_0) & = &\Theta(T-t_0-t), \nonumber
\end{eqnarray}
for positive arguments $t$ and $l$, respectively. Note that the
asymmetrical attribution of the factor $\alpha$ to $\mu_l$ is
arbitrary.

This \emph{ansatz} gives $\sigma = (T-t_0)$ and $\lambda =\alpha
L$. The relations between original and canonical coordinates are
\begin{equation}  \xi(l) = \left\{\begin{array}{lll}
 2\alpha l \qquad && 0<l<L/2\\
 \alpha L\qquad && l>L/2\;,\end{array} \right.
\end{equation}
\begin{equation}  \tau(t) = \left\{\begin{array}{lll}
 t\quad && 0<t<T-t_0 \\
 T-t_0\quad && t>T-t_0\;.\end{array} \right.
\end{equation}

The recurrence PDFs are obtained by inserting this into
Eqs.~(\ref{pll},\ref{ptt}) and transforming back to the original
coordinates. The \emph{average} distributions of distances between
recurrences and reference points are obtained by averaging over
$t_0$. The final results are
\begin{eqnarray}
 \langle p_l(l)\rangle_{t_0} & = & {1\over l} \left[1-{1-e^{-2\alpha lT}\over 2\alpha lT}
\right] \Theta(L/2-l), \label{l-periodic}\\
 \langle p_t(t)\rangle_{t_0} & = & {1\over t} (1-e^{-\alpha Lt}) (1-{t\over T})
 \Theta(T-t),
\label{t-periodic}
\end{eqnarray}

These detailed results are included in order to demonstrate that
exact calculations are possible in the most simple case. But in
more realistic cases no exact results can be expected. As a
general rule, the simple power laws of Eqs.~(\ref{pl},\ref{pt})
will hold for intermediate values of $l$ and $t$, but corrections
will be necessary both for large and for small $l$ and $t$ -- as
follows from Eqs.~(\ref{pll},\ref{ptt}). The corrections render
the distributions finite at small values of the arguments, and
they cut them off at large ones.

The cut-offs at large $l$ and $t$ occur just at the sizes of the
system. Their detailed shapes depend, as suggested by comparing
Eqs.~(\ref{l-periodic},\ref{t-periodic}) with
Eqs.~(\ref{model_p_l},\ref{model_p_t}), on the specific properties
of the system at large scales. The behavior at small distances is
more general.

To see this, let us consider Eq.~(\ref{t-periodic}) in more
detail. There the deviation from the infinite system limit happens
when $\alpha tL\approx 1$, i.e. at a time
\begin{equation}
t \approx t^*(L) \equiv (\alpha L)^{-1}\quad ,\label{t-cutoff}
\end{equation}
which exactly coincides with Eq.~(\ref{t-cutoff-D}) for the
translational invariant case. Since $\alpha$ is the density of
events in space-time, $t^*$ is the average time delay between
successive events. Obviously, \emph{recurrences} cannot follow
each other faster than \emph{events}. Similarly in
Eq.~(\ref{l-periodic}), the deviation from the infinite system
limit happens when $2 \alpha l T \approx 1$, which coincides with
the expression for $l^*(T)$ in the translational invariant case
given by Eq.~(\ref{l-cutoff-D}).

Not only is the scaling of $l^*$ and $t^*$ identical to the
translationally invariant case but also the qualitative behaviors
of $\langle p_l(l)\rangle_{t_0}$ and of $\langle
p_t(t)\rangle_{t_0}$ for $l \ll l^*$ and $t \ll t^*$,
respectively, are identical. This strongly suggests that the
results given in
Eqs.~(\ref{model_p_l},\ref{model_p_t},\ref{l-cutoff-D},\ref{t-cutoff-D})
capture the essential behavior for scales smaller than the large
scale cut-off -- even when translational invariance is explicitly
broken.

\subsection{Correlations between recurrences and properties of recurrences
 with fixed rank}

Let $p(l,t; l',t')$ be the PDF that two events at space-time
positions $(l,t)$ and $(l',t')$ are both records -- not
necessarily subsequent ones. Referring to
Fig.~\ref{recurrence-chain}, and using Bayes' theorem in canonical
coordinates, we are interested in the probability that no other
event occurs in either of the two rectangles associated to the
events, which is determined by the union of the two rectangular
areas. Hence if $\tau>\tau'$ and $\xi<\xi'$, then
\begin{equation}
p(\xi,\tau;\xi',\tau') = e^{-\xi\tau-(\xi'-\xi)\tau'} \Theta(\sigma -
 \tau)\Theta(\lambda - \xi')\;.
\end{equation}
This directly determines $p(l,t; l',t')$.

Integrating over $\xi$ and $\xi'$ gives the joint PDF for having
recurrences at times $\tau$ and $\tau'$,
\begin{eqnarray}
 p(\tau,\tau') & = &\int_0^\lambda d\xi'\int_0^{\xi'} d\xi\;
 e^{-\xi\tau-(\xi'-\xi)\tau'} \Theta(\sigma - \tau)
 \nonumber \\
& = & {1\over \tau\tau'}\left[1 - {\tau e^{-\lambda \tau'}-
 \tau'e^{-\lambda \tau} \over \tau-\tau'}\right]\Theta(\sigma
 - \tau)\;.
\end{eqnarray}
For $\lambda=\infty$, this gives $p(\tau,\tau') =
(\tau\tau')^{-1}$, for $\tau' < \tau < \sigma$ so the two
recurrences are uncorrelated. For finite $\lambda$, records are
correlated; i.e. $p(\tau,\tau') \neq p_\tau(\tau) p_\tau(\tau')$.
For a stationary process, these results hold in the original
coordinate $t$ as well.

Alternatively, let $q(l,t; l',t')$ be the probability that two
events at $(l,t)$ and $(l',t')$ are \emph{successive} records.
Assuming again that $\tau>\tau'$ and $\xi<\xi'$, we now demand
that both are records, as above, and also that no other
\emph{event} happens in the rectangle
$[\xi',\xi]\times[\tau',\tau]$, or
\begin{equation}
q(\xi,\tau;\xi',\tau') = e^{-\xi'\tau}\Theta(\lambda - \xi')\Theta(\sigma
 - \tau)\;.
\end{equation}
Integrating over $\xi$ and $\xi'$ gives the joint PDF for
having successive records at times $\tau$ and $\tau'$ to be
\begin{eqnarray}
q(\tau,\tau') & = &\int_0^\lambda d\xi'\int_0^{\xi'} d\xi\;
e^{-\xi'\tau} \Theta(\sigma - \tau)\\
& = & {1\over \tau^2} \left[1 - (1+\lambda\tau)
e^{-\lambda\tau}\right] \mbox{ for } \tau'< \tau < \sigma \;.
\nonumber
\end{eqnarray}
Hence, times to successive recurrences are always correlated. When
$\lambda=\infty$, the joint PDF is $q(\tau,\tau') =1/\tau^2$ for
$\tau' < \tau < \sigma$.

For the PDF of the ratio of the times of successive records
$x=\tau'/\tau>0$, it directly follows for finite $\lambda$ that
\begin{eqnarray}
  q_\tau(x) &=&  \int_0^{\sigma x} d\tau' q(\tau(x),\tau') \left|\frac{d\tau}{dx}\right| \nonumber \\
   &=& \Theta(1-x) \sum_{i=1}^\infty \frac{(i-1)(-\lambda \sigma)^i}{i \cdot i!}
   , \label{equ:qx}
\end{eqnarray}
which is constant in the interval $[0;1]$. This is also the
result in the original
coordinates, if the system is stationary -- in which case
$t\propto \tau$ and $x=t'/t$.

We now discuss spatial distance distributions of recurrences with
fixed rank $i$, and first consider a finite stationary system
infinitely extended in time $(\sigma = \infty)$. Let
$p^{(i)}_\xi(\xi)$ be the spatial distance PDF for the $i$-th
recurrence following Event-0. For any $i\geq 2$, the recursion
relation
\begin{equation}
p^{(i)}_\xi(\xi) = \int_\xi^{\lambda} d\xi'\;q(\xi|\xi') p^{(i-1)}_\xi(\xi')\;,
 \label{recurs}
\end{equation}
exists. The quantity $q(\xi|\xi')$ is the conditional PDF, given
that the previous recurrence happened at distance $\xi'$, for the
distance of the next recurrence. One easily shows that
\begin{equation}
q(\xi|\xi') = {1\over \xi'} \Theta(\xi'-\xi)
 \label{null_ratio_canon}
\end{equation}
independently of $i$, so that
\begin{equation}
p^{(i)}_\xi(\xi) = \int_\xi^{\lambda} {d\xi'\over \xi'} p^{(i-1)}_\xi(\xi')\;.
\end{equation}
The solution for finite $\lambda$ is
\begin{equation}
p^{(i)}_\xi(\xi) = {\Theta(\lambda -\xi)\over (i-1)!}
\left(\ln{\lambda \over \xi}\right)^{i-1} \;. \label{pil}
\end{equation}
If the event density in original coordinates was $\mu(l) = aD
l^{D-1} \Theta(R-l)/R^D$, i.e., confined to a disc with radius
$R$, then the last equation translates into
\begin{equation}
p^{(i)}_l(l) = a \Theta(R-l) {D^i l^{D-1}\over(i-1)!R^D}
\left(\ln{R/l}\right)^{i-1}\;,\label{Pi}
\end{equation}
while Eq.~(\ref{null_ratio_canon}) gives for the PDF of the ratio
$x=l/l'>0$
\begin{equation}
q_l(x) = D x^{D-1} \Theta(1-x)\;.\label{null_ratio}
\end{equation}

These last results have to be modified when $\sigma<\infty$, i.e.
when there is a finite observation window in time. In that case we
are not guaranteed that at least $i$ recurrences exist, and thus
$p^{(i)}_l(l)$ has to be replaced by the conditional PDF,
conditioned on the existence of $\geq i$ recurrences. That
requires a more extensive development than we take up here.

\subsection{Distribution of the number of recurrences -- or the degree distributions}
\label{sec:network-degree}

The out-degree distribution $P^{out}(k,N)$ is the probability that
a randomly chosen event out of a sequence of $N$ events has $k$
records. This probability can be deduced using previous results
from the theory of
records~\cite{glick78,sibani98,krug05,nevzorov87,nevzorov}. We
assume that the system is stationary, with a finite rate of events
per unit time. We denote the event defining recurrences as
Event-0. We use the fact that recurrences are records in the sense
that each recurrence is an event that is closer to Event-0 than
all previous events that happened after Event-0. Consider a series
of $i$ events following Event-0. The probability that event $j$ is
a record is $1/j$ and the probability that it is not is $(j-1)/j$.
Hence the probability that there is precisely one record in a
series of $i$ events following Event-0 is $P_i(1)= \prod_{j=2}^i
(j-1)/j = 1/i$. Notice that the first event after Event-0 is
always a record. The probability that there are precisely two
records in the series of $i$ events is
\begin{equation}
P_i(2)= \Bigl(\prod_{j=2}^i \frac{j-1}{j}\Bigr) \sum_{l_1=2}^i
\Bigl(\frac{l_1}{l_1 -1} \times \frac{1}{l_1}\Bigr) =
\frac{1}{i}\sum_{l_1=2}^i \frac{1}{l_1 -1} \quad .
\end{equation}
Continuing with standard methods it is possible to show that the
probability of finding precisely $k$ records in a series of $i$
events, $P_i(k)$, is given by
\begin{eqnarray}
P_i(k) &=& \frac{1}{i} \sum_{1<l_1< \cdots <l_{k-1}\leq i} \frac{1}{(l_1-1) \cdots (l_{k-1}-1)}\nonumber \\
 &=& \frac{|S_i^{k}|}{i!}\nonumber\\
 &\approx& \frac{(\ln i)^{k-1}}{i(k-1)!} \quad ,
\end{eqnarray}
where the symbol $S$ indicates Stirling's number of the first kind
and the last expression holds for $i \gg k \gg 1$. Considering
that each event except the last one in the sequence of $N$ events
initiates its own sequence of records, and hence is an Event-0,
gives
\begin{equation}
\label{model_out} P^{out}(k,N) \approx \frac{1}{N}\sum_{i=1}^{N-1}
\frac{(\ln i)^{k-1}}{i(k-1)!} \approx \frac{(\ln (N))^k}{N \; k!}
\quad ,
\end{equation}
where the last step involves approximating the sum as an integral,
which is valid for large $N$. Therefore, the out-degree
distribution for a random process of $N \gg 1$ events is a Poisson
distribution with mean degree $\langle k \rangle \approx \ln
N$~\cite{krug05}.

Furthermore, the probability to have out-degree one,
$P^{out}(1,N)$, can be computed exactly~\cite{glick78,krug05}: For
those nodes the closest event in space is also the closest in
time. For event $i$, this happens with probability $1/(N-i)$.
Thus,
\begin{equation}
\label{p_out_1} P^{out}(1,N) = N^{-1} \sum^{N-1}_{i=1}
\frac{1}{N-i} \approx \frac{\ln(N) + e_M}{N},
\end{equation}
where we have approximated the harmonic series by the
corresponding integral and $e_M \approx 0.58$ is the
Euler-Mascheroni constant. Note that Eq.~(\ref{p_out_1}) is exact
in the limit $N \to \infty$.


For the in-degree distribution, $P^{in}(k,N)$, similar
considerations apply: Event $i$ is a recurrence of event $j$ ($0
\leq j < i$) with probability $1/(i-j)$, which is independent of
$N$. This allows to compute the in-degree distribution
$P_i^{in}(k)$ of event $i$:
\begin{eqnarray}
P_i^{in}(k) &=& \frac{1}{i} \sum^{i-k}_{l_1=0} \, \sum^{i-k+1}_{l_2=l_1+1} \cdots \sum^{i-1}_{l_{k-1}=l_{k-2}+1} \frac{1}{l_1 l_2 \cdots l_{k-1}}\nonumber \\
&=& \frac{|S_{i}^{k}|}{i!} \nonumber\\
&\approx& \frac{(\ln (i))^{k-1}}{i(k-1)!} \quad ,
\label{in_i}
\end{eqnarray}
for $0 < k \leq i-1$ and zero otherwise. Hence
\begin{equation}
P^{in}(k,N) = \frac{1}{N} \sum^{N-1}_{i=1} P_i^{in}(k) \approx
\frac{(\ln (N))^k}{N \; k!}\quad . \label{in}
\end{equation}
As expected for a fully random process, $P^{in}(k,N)$ is
identical to $P^{out}(k,N)$ and well-approximated by a Poisson
distribution with mean degree $\langle k \rangle \approx \ln N$
for $N\gg1$.

\subsection{Degree correlations}

Due to the acausal nature of the null model, the joint probability
$P_i(k^{in},k^{out})$ that event $i$ has in-degree $k^{in}$ and
out-degree $k^{out}$ factors for all nodes $i$. As a result
\begin{eqnarray}
P_i(k^{in},k^{out}) &=& P_{N-i}(k^{out}) P^{in}_i(k^{in}) \\
 &\approx& \frac{(\ln{(N-i)})^{k^{out}-1}}{(N-i)(k^{out}-1)!} \frac{(\ln{(i)})^{k^{in}-1}}{i(k^{in}-1)!}
\nonumber
\end{eqnarray}
This allows us to compute the mean out-degree of all events with a
given in-degree in a sequence of $N$ events
\begin{widetext}
\begin{eqnarray}
\langle k^{out} \rangle(k^{in},N) = \sum_{k^{out}=0}^{N-1}
k^{out} \frac{1}{N} \sum_{i=1}^{N-1} P_i(k^{in},k^{out}) /
P^{in}(k^{in},N)
 \approx 1+ \frac{1}{(\ln{(N-1)})^{k^{in}}} \int_1^{N-1}
 \frac{(\ln{(i)})^{k^{in}}}{N-1-i} d i \quad .
 \label{model_out_in}
\end{eqnarray}
\end{widetext}
The out-degree $\langle k^{out} \rangle$ weakly depends on
$k^{in}$ due to the fact that the rank of each event implicitly
couples its in- and out-degree in a finite sequence of events. For
instance, if the rank of an event is small (large) compared to
$N$, the in-degree is more likely to be small, but the out-degree
is more likely to be large. Consequently, $\langle k^{out}
\rangle(k^{in},N)$ decreases with $k^{in}$ for fixed $N$. For
similar reasons, weak correlations also appear between the
in-/out-degree of a node and the in-/out-degree of its
recurrences. For example, a large (small) in-degree for a node
implies on average a small (large) out-degree for its recurrences.
Similarly, the out-degree (in-degree) of recurrences increases on
average with the out-degree (in-degree) of their Event-0.

\section{Application to Seismic Patterns}

Seismicity is a prime example where localized events in space and
time can be accurately and, with certain caveats, exhaustively
recorded. It is also a phenomenon where the causal features of the
dynamics responsible for the patterns are subject to ongoing
debate and uncertainty. Seismic data involving many earthquakes
occurring over large regions of space and time exhibit a number of
regularities. These include clustering, fault traces and epicenter
locations with fractal statistics, as well as scaling laws like
the Omori and Gutenberg-Richter (GR) laws (see e.g. Refs.
\cite{turcotte,stein,rundle03} for a review).
Given that the associated earthquake
patterns in space and time are readily observable, approaches
based on the concept of spatiotemporal point processes have been
amply demonstrated to be feasible
\cite{bak02,corral03,corral04,davidsen04,davidsen05m,baiesi05}.
In that case, the description
of seismicity is reduced to recording the size or magnitude of
each earthquake, its epicenter and its time of occurrence.

\begin{figure}[htbp]
\includegraphics*[width=\columnwidth]{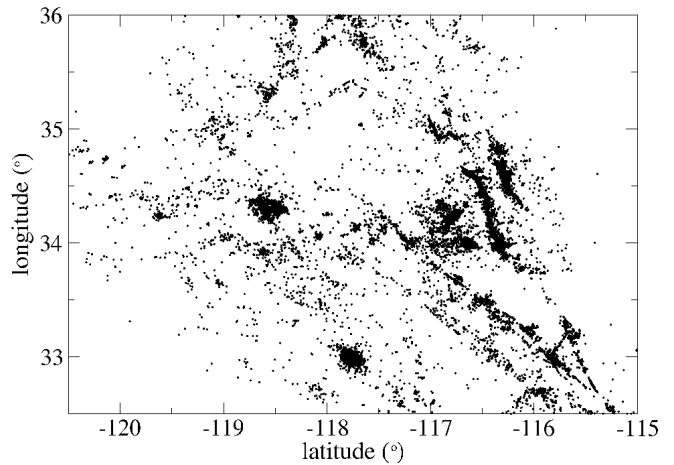}
\caption{\label{map} (Color online) Spatial pattern of seismicity
in Southern California~\cite{catalog}, as described in the text. }
\end{figure}

To test the suitability of our method to characterize seismicity
in a way that makes it possible to infer relevant causal features
of its dynamics and to extend our earlier analysis
\cite{davidsen05pm}, we study a ``relocated'' earthquake catalog
from Southern California \cite{catalog}. The catalog has improved
relative location accuracy within clusters of similar events, the
estimated horizontal standard errors being typically less than 50
to 100m and the estimated vertical standard errors being typically
less than 100 to 200m \cite{shearer03,shearer05}. Due to the
higher relative and absolute location errors for the depth of an
earthquake, we only consider epicenters in the following. The
catalog is assumed to be homogeneous from January 1984 to December
2002 and complete for events with magnitude larger than $m_c=2.5$
located within the rectangle $(120.5^\circ W, 115.0^\circ W)\times
(32.5^\circ N, 36.0^\circ N)$ \cite{wiemer00}. Restricting
ourselves to magnitudes larger than $m_c$ gives $N = 22217$ events
(see Fig.~\ref{map}). In order to test for robustness and the
dependence on magnitude, we analyze this sub-catalog and subsets
of it, obtained in two different ways: By (a) selecting different
threshold magnitudes, namely $m=3.0, 3.5, 4.0$ giving $N=5857,
1770$ and $577$ events, respectively, or (b) using a shorter
period from January 1984 to December 1987 giving $N=4744$ events
for magnitude threshold $m=m_c$.

\begin{figure}[htbp]
\includegraphics*[width=\columnwidth]{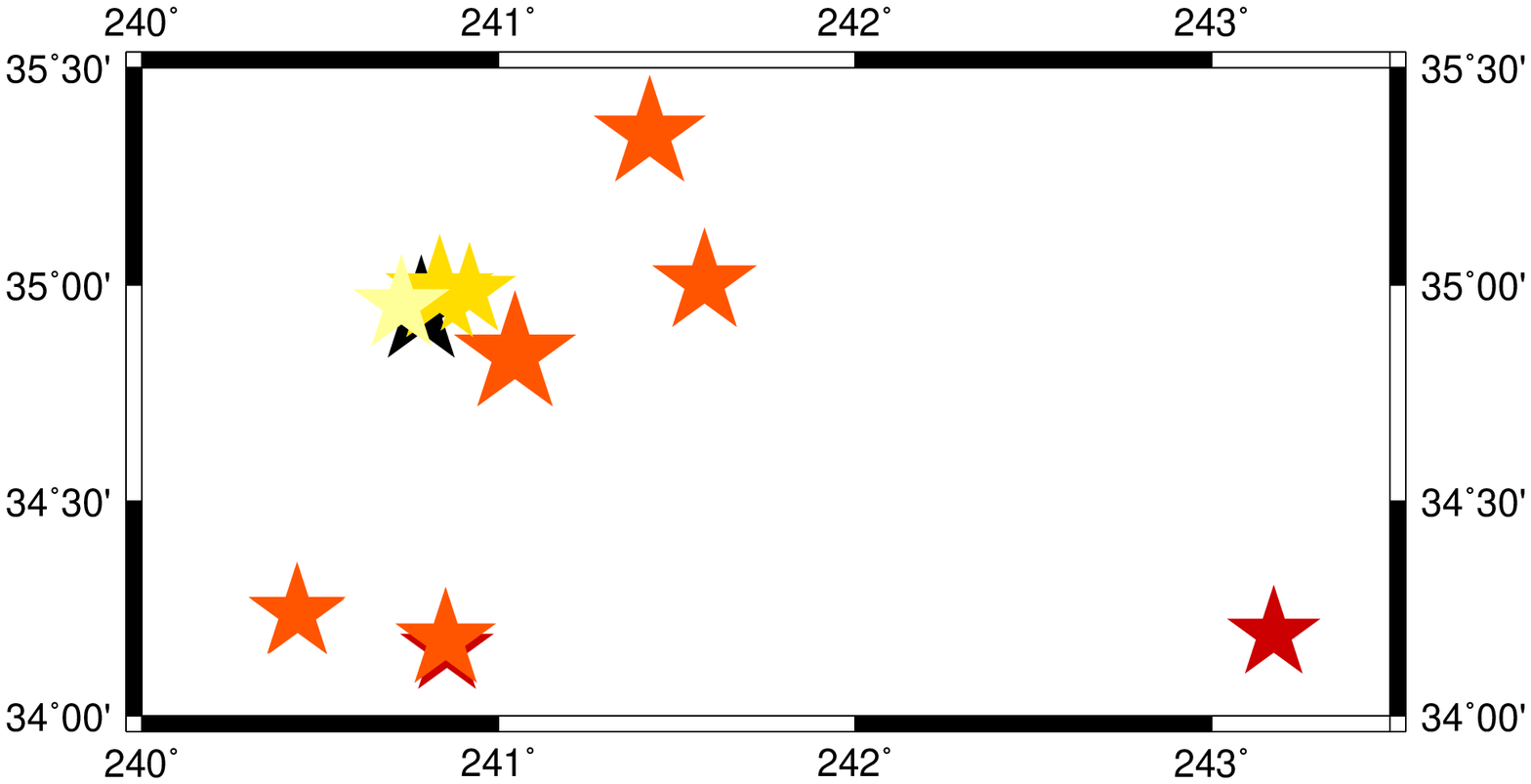}
\caption{\label{hectormine} (Color online) Map showing a 2.9
earthquake (black) and its recurrences as defined by our
method. The size of the symbols linearly scales with the magnitude
of the event and its color corresponds to the time of occurrence
(from darker colors to lighter colors). See
Table~\ref{table_hector} for more details. }
\end{figure}

It is important to note that all events in the catalog are treated
in the same way. In particular, we do not distinguish between
foreshocks, mainshocks and aftershocks. Hence, our definition of a
recurrence -- an event is a recurrence of any previous event if it
is closer to it in space than all the intervening events -- is
\emph{a priori} independent of those classifications. Note also
that our definition of a recurrence is wholly unrelated to the
notion of ``characteristic earthquakes'' on a single fault as
introduced, for example, in
Refs.~\cite{schwartz84,stirling96,matthews02}.

Fig.~\ref{hectormine} shows the recurrences with magnitude $m \geq
m_c$ defined by our method for one randomly chosen event in the
catalog, an earthquake of magnitude 2.9 that occurred
on January 10, 1999. The actual spatial and temporal distance
between this event and each of its recurrences is
listed in Table~\ref{table_hector}. It has to be noted that 
the number of recurrences of a given earthquake 
or Event-0 is generally not related to its magnitude. 
The number of recurrences of the largest earthquakes like the 
Landers event or the Hector mine event are just above the average 
(see Section~\ref{sec:network}).
Thus, most recurrences are associated to Event-0s with small 
magnitude --- which are much more abundant according to the 
Gutenberg-Richter law.  

\begin{table}
\caption{\label{table_hector}List of recurrences of the 2.9
earthquake given in Fig.~\ref{hectormine} as defined by our 
method for threshold magnitude $m=2.5$. }
\begin{ruledtabular}
\begin{tabular}{rcrr}
rank & magnitude & $l$ (km) & $T$ (h)\\
\hline
1 & 2.5 & 234.36 & 22.16\\
2 & 2.5 & 87.39 & 42.81\\
3 & 2.7 & 84.98 & 198.87\\
4 & 2.5 & 84.34 & 232.94\\
5 & 2.6 & 83.97 & 236.56\\
6 & 3.0 & 73.99 & 296.51\\
7 & 2.8 & 72.80 & 424.95\\
8 & 3.3 & 26.37 & 961.64\\
9 & 2.5 & 13.38 & 3471.73\\
10 & 2.9 & 6.99 & 3482.97\\
11 & 2.6 & 5.31 & 25452.30\\
\end{tabular}
\end{ruledtabular}
\end{table}

\subsection{Spatial distances of recurrences}

%
%

\begin{figure}
\noindent\includegraphics*[width=20pc]{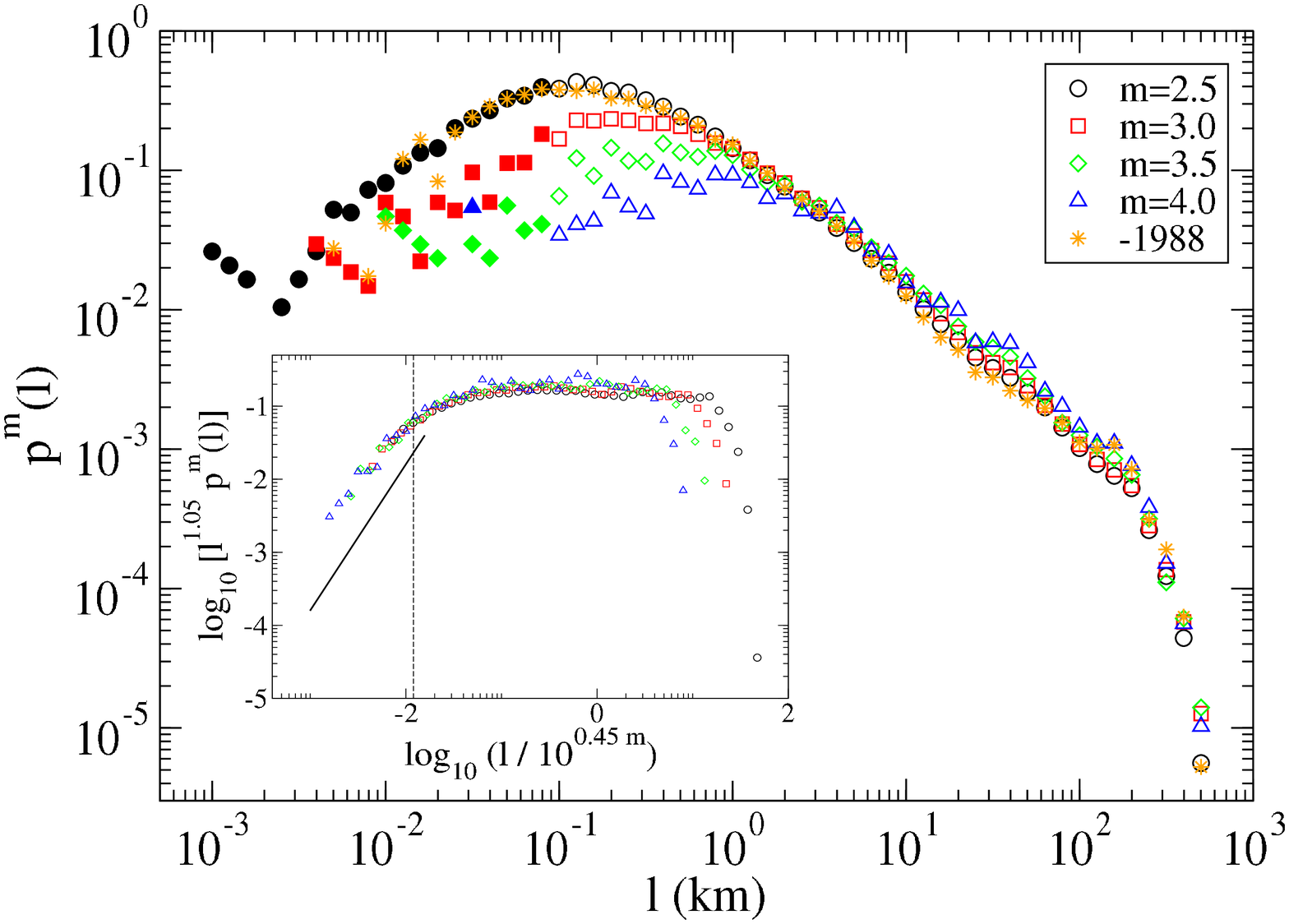}
\caption{\label{fig:l} (Color online) Distribution of distances
$l$ of recurrent events for sets with different magnitude
thresholds $m$. The distribution for $m=2.5$ up to 1988 is also
shown and is almost indistinguishable from the data for the full
catalog with $m=2.5$ -- showing the invariance of the distribution
with respect to the time span of the recorded events. Filled
symbols correspond to distances below 100~m and are unreliable due
to location errors. The inset shows a data collapse, obtained by
rescaling distances and distributions according to
Eq.~(\ref{equ:p(l)})  (excluding unreliable points). 
The full straight line has slope 2.05; the
vertical dashed line indicates the pre-factor $L_0$ in the scaling
law for the characteristic distance, $l^*(m) = L_0 \times
10^{0.45m}$. Note that $\int_0^\infty p^m(l) dl =1$.}
\end{figure}

Fig.~\ref{fig:l} shows the estimated PDF $p^m(l)$ of recurrences
at a spatial distance $l$ in the sub-catalog with threshold
magnitude $m$. The PDFs exhibit a peak at a typical distance,
$l^*(m)$, which increases with magnitude. For sufficiently large
$l$, all distributions show a power law decay with an exponent
$\approx 1.05$ up to a cutoff. This cutoff corresponds to the size
of the region in Southern California that we consider, and hence
is a finite size effect. For small distances $l<l^*(m)$, we
observe an approximately linear increase.

With a suitable scaling ansatz, the different curves in
Fig.~\ref{fig:l} fall onto a universal curve, except at the finite
size cutoff. The inset in Fig.~\ref{fig:l} shows results of a data
collapse using
\begin{equation}
\label{equ:p(l)}
 p^m(l) \approx
 l^{-1.05}F(l/10^{0.45m}) \; .
\end{equation}
The scaling function $F$ has two regimes, a power-law increase
with exponent $\approx 2.05$ for small arguments and a constant
regime at large arguments. The transition point between the two
regimes can be estimated by extrapolating them and selecting the
intersection point, giving $L_0 = 0.012$km. For the characteristic
distance that appears in $F$ we find
\begin{equation}
\label{l_*}
l^*(m) \approx L_0 \times 10^{0.45m}.
\end{equation}

\subsubsection{Discovery of Causal Structure}

Although $p^m(l)$ has the same overall shape as the distribution
$p(l)$ of the finite null model (see Eq.~(\ref{model_p_l})), there
are fundamental differences with respect to the dependence on the
time span over which events are recorded. For the earthquake data,
$p^m(l)$ and in particular $l^*(m)$ do not depend on the time span
at all but rather depend directly on $m$. This conclusion comes
from the explicit comparison of two different observation periods
in Fig.~\ref{fig:l} with the same $m$. With the exception of the
smallest values of $l$, $p^{2.5}(l)$ is largely unaltered if only
the sub-catalog up to 1988 is analyzed and $l^*$ does not change
at all. It is important to note that the total number of events in 
the latter sub-catalog is roughly 5 times smaller.

In the null model $l^*$ depends explicitly on the
finite time span of the observation period, $T$, as shown in
Eq.~(\ref{l-cutoff-D}). In the real data though, the spatiotemporal 
ordering of earthquakes determines the value of $l^*$, regardless of 
the duration of the observation period -- as long as it is large 
enough to obtain sufficient statistics to determine $l^*(m)$ and
small enough that seismic correlations do not disappear over that
time span. This is confirmed by analyses of other sub-catalogs (not 
shown). On this basis, we conclude that the characteristic length 
must therefore reflect robust
physical properties of the underlying dynamics over the given
observation periods. Its (quasi)-invariance is not a property of
the null model. Therefore, it reflects causal structure in the
dynamics of seismicity. As a result, if one re-arranges the
seismic catalog by ``shuffling'' the locations and magnitudes of
events (see Section \ref{sec:shuffling}), then the invariance of
$l^*$ is lost and the distribution of recurrences behaves the same
as the null model for spatial dimension $D=2$ (see
Eqs.~(\ref{model_p_l},\ref{l-cutoff-D})). To sum up: the
invariance of $l^*(m)$ is an indicator of causality and is thereby
a physically meaningful length scale in the dynamics of seismicity
over the time scales we can explore with statistical methods --
minutes to decades.

\subsubsection{Identification with the Rupture Length}

The almost complete lack of dependence of $p^m(l)$ (excluding very
small values of $l$) on the considered time span can be explained 
by at least two scenarios: 1) Recurrences with $l \ll l^*(m)$ are greatly
\emph{suppressed} at large time scales; 2) Recurrences with $l
\approx l^*(m)$ are greatly \emph{enhanced} at short time scales
compared to the null model with \emph{constant} rate. As we will
discuss below, it is likely that both effects are present.

Physically, such a behavior is reasonable if we identify $l^*$
with the rupture length of the earthquake that starts a chain of
recurrences. As described by Omori's law \cite{omori}, the rate of
seismic activity tends to increase directly after an earthquake
nearby (close to the rupture area of the event). Moreover, there is 
some evidence that due to the stress relief within the rupture area 
itself,  it tends to exhibit less seismic activity for 
awhile --- see, for example, Ref.~\cite{rubin00}. This supports the
hypothesis that activity increases for $l \approx l^*(m)$ at shorter 
times, but gets suppressed for $l \ll l^*(m)$ over longer times. 

This identification is also affirmed by the fact that the scaling
of $l^*(m)$ with $m$ is close to the estimated behavior of the
rupture length $L_R(m') \approx 0.02 \times 10^{m'/2}$km given in
Ref.~\cite{kagan02} and remarkably close to $L_R(m') = \sqrt{A_R}
\approx 0.018 \times 10^{0.46\;m'}$km given in
Ref.~\cite{wells94}, where $m'$ is the magnitude of the earthquake
and $A_R$ its rupture area. The close agreement between the latter
and Eq.~(\ref{l_*}) suggests that the characteristic length scale
of distances for recurrent events is indeed the rupture length of
events with $m'=m$, defined in terms of the rupture area $l^* =
L_R \equiv \sqrt{A_R}$. Thus, our approach allows us to discover
the rupture length as a causal consequence of the dynamics based
purely on the spatiotemporal organization of seismicity without
any additional knowledge of the microscopic dynamics and the
actual rupture processes that occur -- even, in fact, treating the
seismic events as point-like in space and time!

The identification $l^*=L_R$ is also consistent with the fact that
the description of earthquakes as a point process breaks down below
the rupture length. Then, the relevant distance(s)
between earthquakes is not determined solely by their epicenter
positions but also by the relative orientation and size of the
extended ruptures in 3D space. Thus, we expect to find a different
correlation structure for distances smaller than the rupture
length. In fact, this is precisely what our data show, namely a
linear increase at small distances, $l \ll l^*(m)$ (see the main
part of Fig.~\ref{fig:l} and also the straight line with a slope
of 2.05 in the inset of Fig.~\ref{fig:l}).

\subsubsection{Robustness of $l^*(m)$}

The lengths $l^*$ observed for the values of $m$ we consider are
larger than the location errors ($\approx 100$m). Simulations show
that $p^4(l)$ (blue triangles in Fig.~\ref{fig:l}) does not change
substantially if the epicenters in the catalog are randomly
relocated by a small distance up to one kilometer. Yet, the
maximum for $p^{2.5}(l)$ shifts to larger $l$ with this procedure,
destroying the scaling of $l^*(m)$. Since the smallest $l^*$ that
obeys the data collapse is $\approx 160$m, the data collapse we
observe for the original data verifies that the relative location
errors are indeed less than $100$ m, or of that order \cite{note2}.
Furthermore, the absence of any anomaly due to location errors
near $100$m in Fig.~\ref{fig:l} indicates that recurrences within
the rupture area lack correlations. This is also supported by
Eq.~(\ref{model_p_l}) which predicts the observed behavior $p^m(l)
\propto l$ for $l<l^*$ if events are happening uniformly and
randomly in 2D space during a finite observation period, or are
recorded as happening randomly in space due to location errors.

%
%

\subsubsection{Spatial Hierarchy of Subsequent Records}
\label{sec:hierarchy}

To further examine the behavior of $p^m(l)$, we study separately
the contributions of recurrences with definite rank. The rank $i$
is defined as in Sec. III F, i.e., for a given Event-0, recurrence
$i+1$ directly follows recurrence $i$ as shown in
Fig.~\ref{recurrence-chain}. Since $p^m(l)$ is the PDF that
\emph{any} recurrence occurs at distance $l$ for a catalog with
threshold magnitude $m$, we have for any finite number of events
$N$,
\begin{equation}\label{l_sum}
p^m(l) = \sum_{i=1}^{N-1} p_{rec}(i) p^m_i(l) =
\frac{\sum_{i=1}^{N-1} N_i p^m_i(l)}{\sum_{i=1}^{N-1} N_i},
\end{equation}
where $p_{rec}(i)$ is the probability that a randomly chosen
recurrence is an $i$'th recurrence, $N_i$ is the number of events
in the sequence that have at least $i$ recurrences (or out-going
links), and $p^m_i(l)$ is the conditional PDF that, given that a
recurrence is an $i$'th recurrence, it happens at distance $l$.

\begin{figure}[htbp]
\includegraphics*[width=\columnwidth]{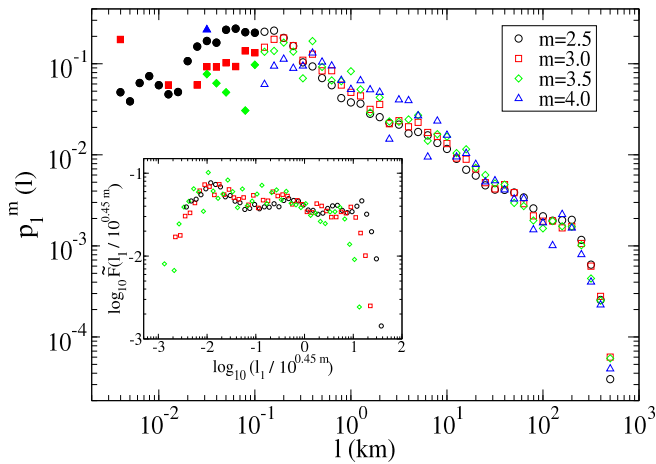}
\caption{\label{rank1} (Color online) Distribution of distances
$l$ of the \emph{first} recurrence for different magnitude
thresholds $m$. Filled symbols correspond to distances below 100~m
and are unreliable due to location errors. Note that
$\int_0^\infty p^m_1(l) dl =1$. Inset: Data collapse obtained by
rescaling distances and distributions according to
Eq.~(\ref{l_1}) (excluding unreliable points).}
\end{figure}

In the inset of Fig.~\ref{rank1}, the data are analyzed according
to the {\it{ansatz}} that the distribution of first recurrences,
$p^m_1(l)$, has the scaling form
\begin{equation}\label{l_1}
p^m_1(l) = l^{-\delta_r} {\tilde F}(l/10^{0.45\;m}),
\end{equation}
with $\delta_r \approx 0.6$ and ${\tilde F}$ similar to $F$ (see
Eq.~(\ref{equ:p(l)}) and the inset of Fig.~\ref{fig:l} for
comparison). In particular, the same characteristic distance
$l^*(m)$ appears as for $p^m(l)$. Moreover, we find that the
latter is true for all $p^m_i(l)$ --- which is further evidence supporting
the interpretation of $l^*$ as the rupture length. The
behavior of $p^m_1(l)$ indicated in Fig.~\ref{rank1} and described
by Eq.~(\ref{l_1}) extends earlier results for a catalog from
Southern California with lower spatial resolution ($\approx 1$km)
which did not allow to resolve the dependence on $m$
\cite{davidsen05m}.

\begin{figure}[htbp]
\includegraphics*[width=\columnwidth]{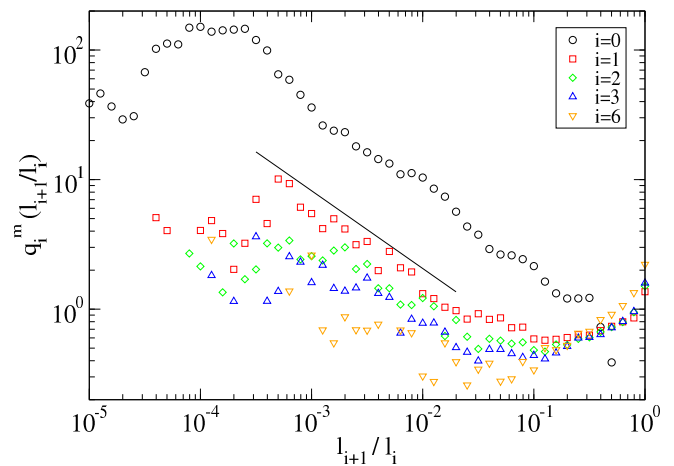}
\caption{\label{rank_ratio_l} (Color online) Distribution of
recurrence distance ratios $l_{i+1}/l_i$ for $m=2.5$ and different
values of $i$ with $l_0 = 634.3$ km. The straight line corresponds
to a decay with exponent 0.6. Note that $\int_0^\infty q^m_i(x) dx
=1$.}
\end{figure}

Related to the distribution of distances for recurrent events is
the distribution of distance ratios $l_{i+1}/l_{i}$ of consecutive
recurrences. Here again recurrences are ordered such that
recurrence $i+1$ directly follows recurrence $i$. For $i=0$, we
take $l_0 = 634.3$ km, which is the largest possible distance in
the region covered by the catalog. By construction these ratios
are always between zero and one. We denote by $q^m_i(x)$ the PDF
that $l_{i+1}/l_{i}=x$ for each event that has an $(i+1)^{th}$
recurrence. As indicated in Fig.~\ref{rank_ratio_l}, the data for
$m=2.5$ and $i=0$ (black circles) scale over a wide region as
$q^{2.5}_0(x)\sim x^{-\delta_r}$ with $\delta_r \approx 0.6$. This
is expected since $q^m_0(x) \sim p^m_1(l)$. Although each
distribution $q^{2.5}_i(x)$ is different, the curves for $i \geq
1$ also show (more restricted) power law decay comparable to
$q^{2.5}_0(x)$. For $l_{i+1}/l_i \to 1$ they also exhibit a peak
that becomes more pronounced with increasing $i$. This is due to
recurrences occurring at almost the same distance (but not at the
same place!) suggesting again that recurrences are suppressed within the
rupture area, but are enhanced just outside that area.


The observed behavior of $q^{2.5}_i(x)$ and $p^m_1(l)$ is very
different from the behavior predicted by the null model. For the
null model, in the long time limit, Eq.~(\ref{null_ratio}) gives
$q^m_i(x) = D x^{D-1}$ which is not only independent of $i$ but
also purely determined by the spatial dimension $D$ -- and is
increasing for $D>1$ rather than decreasing. Similarly,
Eq.~(\ref{Pi}) gives $p^m_1(l) \propto l^{D-1}$ for the null
model. For Southern California, it has been found that $D = D_2 =
1.2$ \cite{davidsen04,davidsen05m}, which would lead to an
increasing function $q^m_i(x)\sim x^{0.2}$ rather than a decaying
power law behavior. Although the above predictions of the null
model are only strictly true in the infinite time limit, we point
out that repeating this analysis of the hierarchy of recurrences
for a ``shuffled'' catalog reveals behavior in close agreement
with the null model and diametrically opposed to the results shown
in Fig.~\ref{rank_ratio_l} for the actual seismic record
\cite{davidsen04}.

Thus, the observed behavior of $q^{2.5}_i(x)$ and $p^m_1(l)$ as
well as the value of $\delta_r$ are \emph{not} determined by the
spatial distribution of seismicity alone but reflect causal
structures leading to the complex \emph{spatiotemporal}
organization of seismicity. Moreover, the shape of $p^m_1(l)$
shows that the first recurrence is much more likely to happen at a
typical distance of $l^*$ than predicted by the null model. This
enhancement goes along with a suppression of recurrences with $l
\ll l^*$ as the increasing (with $i$) peak at $x=1$ for $q^m_i(x)$
indicates. These results support the overall picture that
recurrences with $l \ll l^*(m)$ are greatly \emph{suppressed} at
large time scales while recurrences with $l \approx l^*(m)$ are
greatly \emph{enhanced} at short time scales.

\subsection{Network properties\label{sec:network}}

%
%

\begin{figure}[htbp]
\includegraphics*[width=\columnwidth]{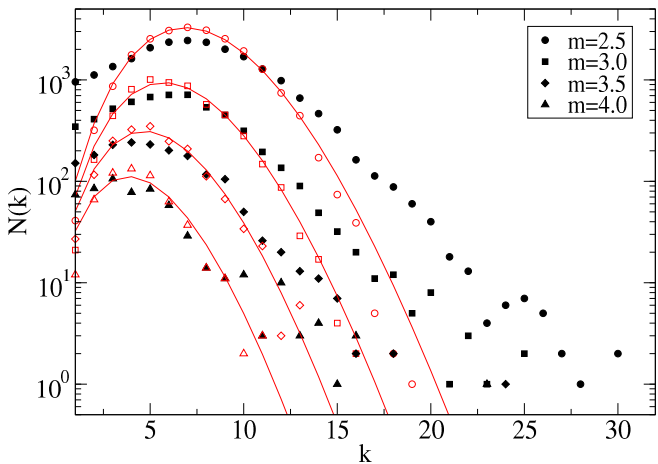}
\caption{\label{in_out} (Color online) In- and out-degree
histograms for different values of $m$. For a given earthquake,
the in-degree (out-degree) $k$ is the number of links directed at
it (originating from it) as defined in Section~\ref{sec:synth}.
Open (red) symbols correspond to the in-degree, filled (black)
symbols correspond to the out-degree. Error bars can be estimated
as $\sqrt{N(k)}$. The red lines correspond to Poisson
distributions with the same respective mean and normalization.}
\end{figure}

We now turn to the analysis of seismicity in terms of the
statistical properties of its network of recurrences (or records)
as defined in Section~\ref{sec:synth} and illustrated in
Fig.~\ref{cascade}. Fig.~\ref{in_out} shows the in- and out-degree
histograms for different values of $m$, which are compared to
Poisson distributions with the same respective mean degree and
normalization ($\langle k \rangle = 7.40, 6.24, 5.20, 4.35$ for $m
= 2.5, 3.0, 3.5, 4.0$, respectively). A Poisson out-degree and
in-degree distribution is expected for the null model (see
Eq.~(\ref{model_out}) and Eq.~(\ref{in})). For the actual seismic
network, the out-degree distributions are significantly different
from a Poissonian \cite{note3}. In particular, the network keeps a
preponderance of nodes with small out-degree as well as an excess
of nodes with large out-degree compared to a Poisson distribution.
This effect becomes more pronounced with increasing magnitude.

The behavior of the out-degree distribution implies that the
network topology is able to discern consequences of the causal
structure of seismicity: The preponderance of nodes with small
out-degree, for example, can be related to the physical picture
discussed above that seismic activity is typically greatly
enhanced directly after the occurrence of an earthquake close to
its rupture area but suppressed within the rupture area itself.
Such a dynamics makes it more likely that only very few
recurrences occur, even at long times. For the in-degree
distributions, we find that they roughly agree with a Poisson
distribution although there are still significant deviations from
the null model for $k=1$ \cite{note4}.

Note, however, that $\langle k \rangle$ --- which is obviously the same for
the in- and out-degrees --- decreases with $m$, simply because the
number of events $N$ shrinks with $m$. This is shown in
Fig.~\ref{mean_k} where $\langle k \rangle$ is also displayed for
a randomly shuffled catalog.

\subsubsection{Shuffling Procedure}
\label{sec:shuffling}

\begin{figure}[htbp]
\includegraphics*[width=\columnwidth]{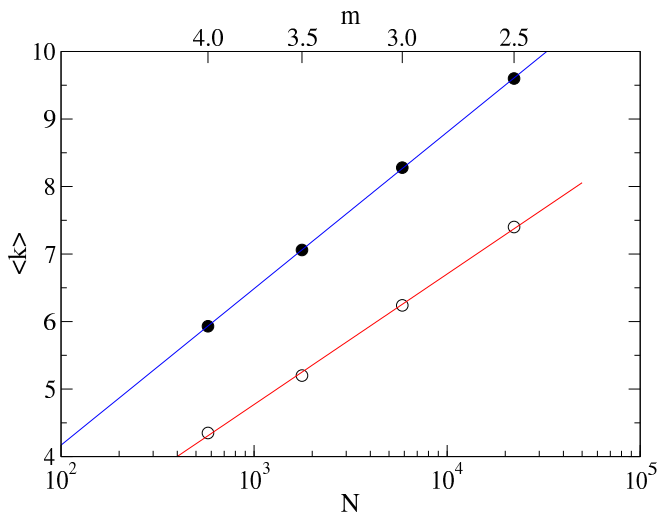}
\caption{\label{mean_k} (Color online) Mean degree $\langle k \rangle$ as a
function of number of events $N$ (or magnitude $m$). Open symbols
correspond to the original catalog for different values of $m$,
while filled symbols correspond to the shuffled catalog (see
text). The lines correspond to best fits giving $\langle k
\rangle_{original} = -1.03 + 0.84\;\ln{N}$ and $\langle k
\rangle_{shuffled} = -0.47 + 1.01\;\ln{N}$.}
\end{figure}

Shuffling was performed in the following way: Consider all events
in the catalog with magnitude $m'\geq m_c=2.5$. Shuffle the
magnitudes and the epicenter locations separately, keeping the
times of occurrence, and then apply the recurrence analysis for
the different subsets defined by different magnitude thresholds as
before. The shuffled catalog can, thus, be considered as a
realization of a random process with no spatiotemporal
correlations, although both spatial correlations and temporal
correlations may persist separately. Based on the null model and
Eq.~(\ref{model_out}), we expect a Poisson out-degree distribution
with $\langle k \rangle \approx \ln{N}$ which is exactly what we
find for the randomly shuffled catalog.
%
%
This dependence of $\langle k \rangle$ can be clearly seen in
Fig.~\ref{mean_k}. Yet, for the original earthquake data we find
for large $N$
\begin{equation}
\langle k \rangle \approx 0.8 \; \ln{N}.
\end{equation}
Hence, the average number of recurrences is significantly less
than for the null model, which is presumably related to the suppression
of recurrences with $l<l^*(m)$ -- as discussed
earlier. Fig.~\ref{mean_k} gives further evidence that recurrences
emphasize particular aspects of spatiotemporal clustering,
associated with the causal dynamics of seismicity.

\subsubsection{Degree-degree Correlations}
\begin{figure}[htbp]
\includegraphics*[width=\columnwidth]{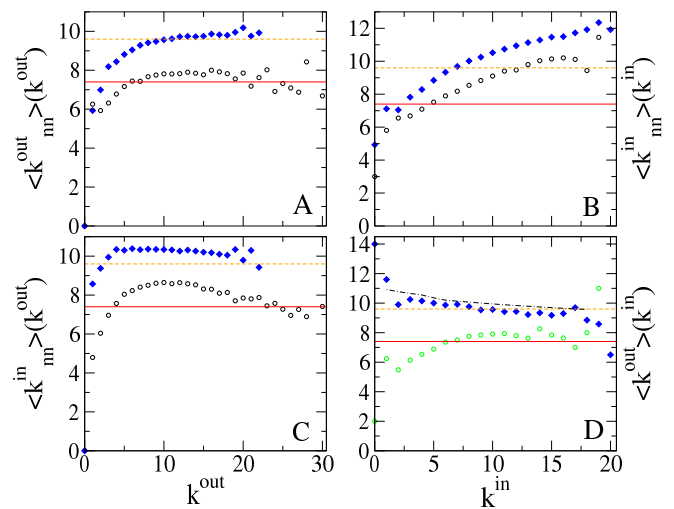}
\caption{\label{fig:dd} (Color online) Panels A -- C: degree
correlations between the Event-0 and its recurrences. The average
in-degree $\langle k^{in}_{nn} \rangle$ or out-degree $\langle
k^{out}_{nn} \rangle$ of the recurrences of all nodes with a given
in-degree $k^{in}$ or out-degree $k^{out}$ for $m=2.5$ is shown.
Open (black) circles correspond to the original earthquake catalog
from California, filled (blue) diamonds correspond to the shuffled
catalog. The mean degree is indicated by the (red) solid line and
the (orange) dashed line, respectively. Panel D shows the average
out-degree of a node as a function of its in-degree. Open (green)
circles correspond to the original earthquake catalog from
California, filled (blue) diamonds correspond to the shuffled
catalog. The black dash-dotted line is the approximation for the
null model given in Eq.~(\ref{model_out_in}). In this panel, the
behavior of the original earthquake data is qualitatively very
different from the null model and the shuffled catalog.}
\end{figure}

The causal structure of seismicity does not, however, induce
strong degree-degree correlations between events and their
recurrences other than those arising from the temporal order of a
finite sequence of events -- as in the acausal null model. Panels
A to C in Fig.~\ref{fig:dd} show the average out-degree and
in-degree of recurrences as a function of the in-degree or
out-degree of their corresponding Event-0 \cite{note5}. There are
no qualitative differences between the actual earthquake catalog
from California and a surrogate, which is a randomly shuffled
version of the catalog. In particular, the behavior shown in panel
A and B agrees with the acausal null model (see discussion
following Eq.~(\ref{model_out_in})). Note that the offset between
the two data sets is simply due to different $\langle k \rangle$.

The situation is different for the dependence of the mean
out-degree on the in-degree of the {\it same node}. As shown in
Eq.~(\ref{model_out_in}), $\langle k^{out} \rangle$ has a weak
dependence on $k^{in}$ in the null model such that $\langle
k^{out} \rangle$ decreases with $k^{in}$. This is exactly what we
find for the shuffled catalog as shown in panel D of
Fig.~\ref{fig:dd}. However, the same panel also shows that for the
actual earthquake catalog $\langle k^{out} \rangle$ increases with
$k^{in}$ -- exactly the opposite of the null model. Moreover,
$k^{in} < \langle k \rangle$ implies $k^{out} < \langle k \rangle$
on average. This is again consistent with a causal dynamics where
earthquakes are clustered in space and time.

%
%
\subsubsection{Clustering coefficient}

Other network properties include various measures of clustering.
In general terms, clustering quantifies how well connected the
neighbors of a node are among themselves. In the case of
recurrences, it refers to the likelihood that recurrences of the
same event are also recurrences of each other. There are
different, inequivalent definitions of the clustering coefficient
$C$~\cite{soffer05}. Here we focus on the definition based on the
local clustering coefficient $C_i$ adapted to directed networks.

For all nodes $i$ with out-degree larger than one, the clustering
coefficient $C_i$ is given by the ratio of existing links $E_i$
between its $k^{out}_i$ recurrences to a possible number of such
links, $\frac{1}{2} k^{out}_i (k^{out}_i-1)$. Then the
clustering coefficient $C$ of the network is defined as the
average over all nodes $i$ with out-degree larger than one
\begin{equation}
C = \langle C_i \rangle = \left \langle \frac{2
E_i}{k^{out}_i(k^{out}_i-1)} \right \rangle_i. \label{eq_c}
\end{equation}
This definition implies, for example, that the clustering
coefficient of an Erd\"os-Renyi graph is equal to the the
probability of linking each pair of nodes, $p_{link}=\langle k
\rangle / (N-1)=C_{rand}$.

For the data from California, we obtain $C=0.2647$ for $m=2.5$.
This is significantly larger than $C=0.1825$, which is the value
for the shuffled catalog. It has to be pointed out, though, that
the average is performed over a different number of nodes in the
two cases since the shuffled catalog hardly contains any events
with out-degree equal to one. For the shuffled catalog, there are
only 15 events with $k^{out}=1$, which is close to the expected
value of $10.6$ for the random model -- see Eq.~(\ref{p_out_1}).
This value is two orders of magnitude less than for the actual
seismic data.

Another difference between the two data sets is the distribution
of $C_i$. For the actual earthquake data, the distribution is much
broader. The standard deviation for the distribution is $0.2146$
compared to $0.0934$ for the shuffled catalog. This difference is
mainly due to the fact that the original data contain many events
with $C_i=0$ or $C_i=1$ -- unlike the shuffled catalog.

\subsection{Temporal distances of recurrences}

%
%

\begin{figure}[htbp]
\includegraphics*[width=\columnwidth]{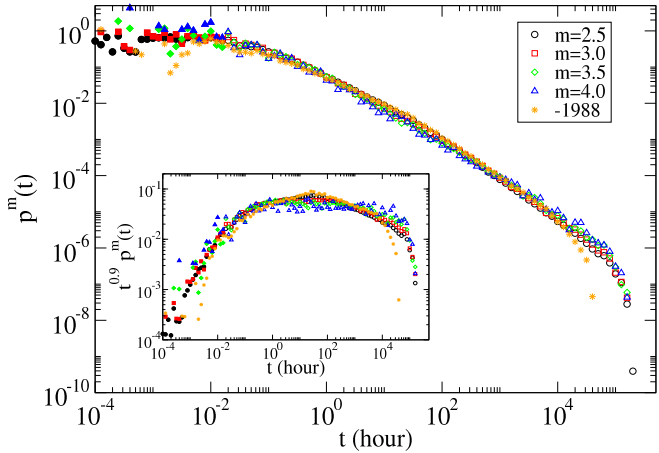}
\caption{\label{fig:t} (Color online) Distributions of the waiting
times between Event-0s and their recurrences for the original
catalog and different threshold magnitudes $m$. The distribution
for $m=2.5$ up to 1988 is also shown. Filled symbols correspond to
times below 90 seconds which are underestimated and unreliable due
to measurement restrictions: The finite rupture times of
earthquakes and the associated seismic coda, which consists of a
superposition of incoherent scattered waves, place limitations on
the identification and separation of earthquakes. The inset shows
the rescaled distributions. Note that $\int_0^\infty p^m(t) dt
=1$.}
\end{figure}

The temporal distances between events and their recurrences can be
analyzed in the same way as the spatial distances. The PDF
$p^m(t)$ for these waiting (or ``inter-occurrence'') times for
different threshold magnitudes $m$ is shown in Fig.~\ref{fig:t}.
These all decay roughly as $1/t^{\alpha}$ with $\alpha\approx 0.9$
for intermediate times as indicated in the inset. The apparent
scaling region in Fig.~\ref{fig:t} shows some curvature, though.
Due to the finite duration of the catalog, there is an
observational cut-off at the longest time scales. At the shortest
time scales, $p^m(t)$ goes over to a constant limit. While the
shape of the distribution is roughly similar to the null model
(see Eq.~(\ref{model_p_t})), $p^m(t)$ for the earthquake catalog
is \emph{independent} of $m$ and, hence, the number of events in
the catalog. This invariance is (again) drastically at odds with
the null model where the temporal rate $\Lambda=a b R^D$
determines the transition point and $\Lambda$ itself depends on
the number of events $N$ as shown in
Eqs.~(\ref{t-cutoff-D},\ref{t_N}).

\begin{figure}[htbp]
\includegraphics*[width=\columnwidth]{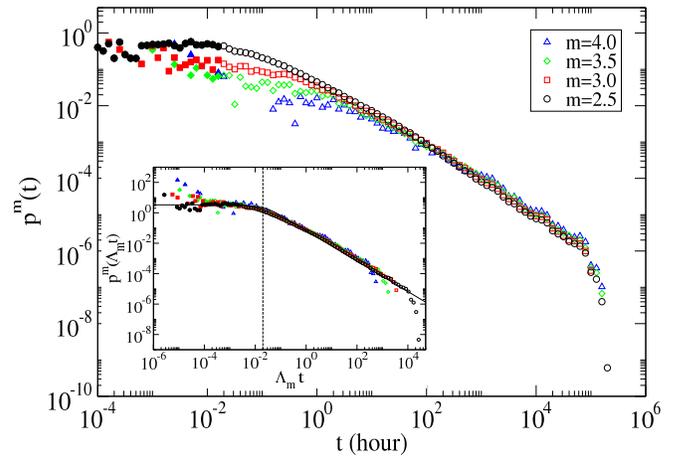}
\caption{\label{fig:shuff_t} (Color online) Distributions of the
waiting times between Event-0s and their recurrences in the
shuffled catalog (see text) for different threshold magnitudes
$m$. Filled symbols correspond to times below 90 seconds which are
underestimated and unreliable. The inset shows the distributions
rescaled by the respective rate of events $\Lambda_m$. The solid
line corresponds to a best fit assuming the functional form given
in Eq.~(\ref{model_p_t}). The dashed line highlights the
transition point between the constant behavior and the $1/t$
decay. Note that $\int_0^\infty p^m(t) dt =1$.}
\end{figure}

As described in what follows, the analysis for the shuffled catalog shown in
Fig.~\ref{fig:shuff_t} is consistent with the acausal null model.
As predicted by the null model, the distributions for the shuffled
catalog must be rescaled by the rate of events in order to obtain
a data collapse. Furthermore, the invariant behavior (with respect
to magnitude $m$) we observe for recurrences in the original
catalog differs substantially from earlier results for waiting
time distributions between subsequent earthquakes
\citep{bak02,corral03,corral04,davidsen04}. It reflects a new
non-trivial feature of the spatiotemporal dynamics of seismicity
that appears when events other than the immediately subsequent
ones -- used to conventionally define waiting times -- are
considered.

For the shuffled catalog, $p^m(t)$ closely follows the theoretical
prediction of Eq.~(\ref{model_p_t}) and in particular the
dependence on $m$ -- or rather on $N$ through $\Lambda$. As shown
in the inset of Fig.~\ref{fig:shuff_t}, the different
distributions --- with the obvious exception of the observational
cut-off at the largest time scales --- collapse onto a single
curve if $t$ is rescaled by the respective rate $\Lambda_m$. Here,
$\Lambda_m$ is the mean rate of earthquakes above magnitude
threshold $m$ for the observation period. Notably, the main
deviation from the \emph{stationary} null model is that the
location of the transition point for the shuffled catalog is not
at $\Lambda_m t =1$ but rather at $\Lambda_m t = 0.02$. This is
expected and due to the fact that the rate of seismic activity ---
which is preserved in the shuffled catalog
--- is \emph{not} constant over time but exhibits large, correlated
fluctuations as indicated, for example, by Omori's law
\cite{omori}.

%
%

\begin{figure}[htbp]
\includegraphics*[width=\columnwidth]{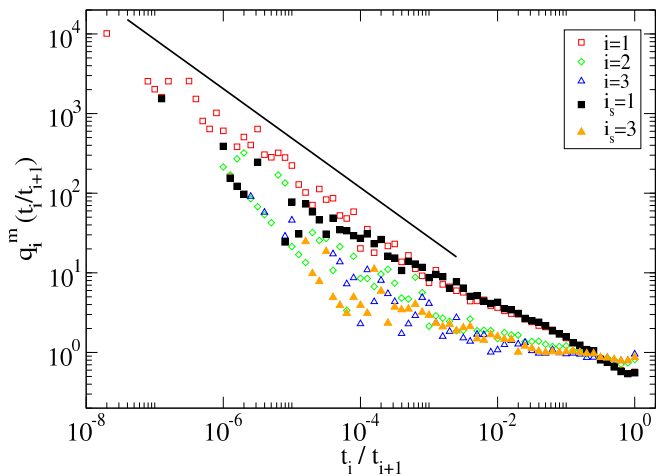}
\caption{\label{fig:ratio_t} (Color online) Distribution of
waiting time ratios $t_{i}/t_{i+1}$ for $m=2.5$. The straight line
has slope -0.62. Open symbols correspond to the original
earthquake catalog, filled symbols to the shuffled catalog (see
text). Note that $\int_0^\infty q^m_i(x) dx =1$.}
\end{figure}

The relative times between subsequent recurrences in the hierarchy
can be analyzed in the same way as distances were in
Sec.~\ref{sec:hierarchy}. Fig.~\ref{fig:ratio_t} shows the PDFs
for the ratios $t_i/t_{i+1}$ for subsequent recurrences, i.e.,
recurrences are ordered such that recurrence $i+1$ directly
follows recurrence $i$. For the cases shown, two power-law regimes
seem to exist: For arguments smaller than about $10^{-3}$,
$q^{2.5}_i(t_{i}/t_{i+1})$ decays with an exponent $\delta_t
\approx 0.6$ roughly independent of $i$, for larger arguments the
decay is slower and the exponent apparently decreases further with
$i$. Clearly, the broadest scaling regime materializes for
$t_1/t_2$.

The behavior of $q^{2.5}_1(t_{1}/t_{2})$ for $10^{-3} <
t_{1}/t_{2} \ll 1$ could be compared to Eq.~(\ref{equ:qx}),
although the latter was derived for the translational
invariant case. Equally important, Eq.~(\ref{equ:qx}) only
holds for the \emph{stationary} null model. As discussed above,
seismic activity is \emph{not} constant over time but exhibits
large fluctuations. Fig.~\ref{fig:ratio_t} shows that these
fluctuations as well as the loss of translational invariance are
responsible for the behavior for arguments larger than about
$10^{-3}$, since there is no observed difference between the
original and the shuffled catalog.
Yet, the deviations between the original data and the
shuffled catalog for smaller arguments indicate that those short
time differences arise
from the causal spatiotemporal organization of seismicity.

\subsection{Discussion}

It is important to discuss our results for the network of
recurrences (or records) in view of what is known about causal
connections between earthquakes. One specific type of causal
connection is earthquake triggering. The increased seismic
activity following large earthquakes --- as described by the Omori
law \cite{omori} leading to the identification of aftershocks ---
is the most obvious example of earthquakes being triggered in part
by preceding events. Aftershock sequences of small earthquakes are
less obvious because the aftershock productivity is weaker, but
can be observed after stacking many sequences \cite{helmstetter05}. 
Other approaches [16, 25] have generalized the definition of an 
aftershock so that an event can be an aftershock of more than one 
event leading to networks of earthquakes and aftershocks. 
Earthquake triggering is typically associated
with stress changes which can be static stress changes imparted by
the preceding shock or dynamic stress changes associated with
seismic wave propagation or combinations of them as discussed, for
example, in
Refs.~\cite{harris98,stein99,freed05,main06,mallman07}. The
proposed physical mechanisms to explain earthquake triggering due
to \emph{static} stress change induced by a prior event include
rate-and-state dependent friction \cite{dietrich94}, crack growth
\cite{das81,shaw93,main00}, viscous relaxation \cite{mikumo79},
static fatigue \cite{scholz68}, pore fluid flow \cite{nur72}, and
simple sandpile models \cite{hergarten02}.

Calculations of stress changes have been used to predict the
locations, focal mechanisms and times of future earthquakes (see
Refs.~\cite{harris98,stein99,king01} for reviews). The success of
this method is limited. Only about 60\% of aftershocks are located
where the stress increased after a main shock \cite{parsons02};
stress shadows are seldom or never observed
\cite{marsan03a,felzer03}; and the correlation of stress change
with aftershocks is rather sensitive to the assumed slip
distribution \cite{steacy04}. All of this could be due to the fact
that most studies have neglected the influence of small
earthquakes and secondary aftershocks which can play an important
role \cite{felzer02,helmstetter05}. Moreover, most studies have
also neglected the influence of \emph{dynamic} stresses radiated
by seismic waves from (small or medium-sized) earthquakes which
may also play an important role --- even in the near field (see,
e.g.,
Refs.~\cite{kilb00,gomberg03,kilb03,parsons05,johnson05,pollitz06}).
In particular, dynamic stress changes can dominate the triggering
mechanism over a wide range of distances between 0.2 and 50
kilometers from the fault rupture \cite{felzer06}.

While it is not entirely clear how our results for the network of
recurrences could allow one to distinguish between the different
types of stress changes associated with earthquake triggering,
there are a number of
currently unexplained observations that could be related to a
particular triggering mechanism. The excess of events with a large
number of recurrences compared to the null model (see
Fig.~\ref{in_out}) is one of them. Other examples include the
correlations between the in-degree and the out-degree of a given
event (see Fig.~\ref{fig:dd}~D) and the apparent invariance of the
waiting time distribution with respect to the threshold magnitude
(see Fig.~\ref{fig:t}). The sensitivity of these properties as well
as our other findings (especially the invariance of $l^*(m)$ with
respect to the time span) to the triggering mechanism
can be tested within the framework of the ``epidemic type aftershock sequence''
model which has been established as an improved stochastic null
model for seismicity \cite{kagan87,ogata88}. It allows one to
vary the spatial scaling of the triggered events depending on
the assumed underlying triggering mechanism, namely static stress
changes or dynamic stress changes \cite{helmstetter06,felzer06}.
This will be the topic of a future publication.

Finally, we would like to point out that simple and direct
comparisons of our results for the network of recurrences (or
records) with known results for aftershocks are not justified.
This is due to the fact that recurrences as defined by our method
are at best a very small and \emph{non-random} subset of what
typically would be considered the set of aftershocks. Also, the
power-law decay of the distribution of distances as shown in
Fig.~\ref{fig:l} occurs generically for a wide class of processes
due to the properties of records as discussed in
Section~\ref{sec:null} --- independent of the specific properties
of aftershock sequences described, for example, in
Ref.~\cite{felzer06}. Similarly, the power-law decay of the
distribution of waiting times (see Fig.~\ref{fig:t}) is also a
generic property of records as discussed in Section~\ref{sec:null}
and is, thus, not related to the specific characteristics of
aftershock sequences discussed in
Refs.~\cite{shcherbakov05,shcherbakov06}.

\section{Summary}

This paper provides a method to detect features in a temporal
sequence of observations that can be plausibly attributed to
causal dynamics even when the observer has no \emph{a priori}
knowledge of the underlying dynamics. Our starting point is to
generalize the concept of a recurrence for a point process in time
to recurrent events in space and time. An event is defined to be a
recurrence of any previous event if it is closer to it in space
than all the intervening events; i.e. if it constitutes a record
breaking event. Hence, the causal structure of events may be
described as a network of events linked to their recurrences. Each
event can have many previous events pointing to it (its potential
causes) and many future events (its effects). Causality can be
plausibly inferred when the statistical properties of the network
constructed using this method and the statistics of the records
deviate strongly from those resulting from almost any acausal process.

We derive analytically many properties for the network of
recurrent events composed by random processes in space and time.
In doing so, we develop a fully symmetric theory of records where
both the variable in which records occur and time, itself, are
continuous. This simplifies the theory and in our view makes it
more elegant. We discover a number of new analytic results for
record breaking statistics.

Many of those results are compared to properties of the network
synthesized from time series of epicenter locations for
earthquakes in Southern California. Significant disparities that
can be attributed to causality are mainly coming from the
invariance of network statistics with the time span of the events
considered. This is presumably related to an observed hierarchy in
the distances and times of subsequent recurrences. As a result a
fundamental length scale for recurrences is obtained solely from
the earthquake epicenter data, which can be identified as the
rupture length. All these significant deviations disappear when
the analysis is repeated for a surrogate in which the original
magnitudes and locations of earthquake epicenters are randomly
``shuffled''. Almost all of the latter results are completely
consistent with predictions from the acausal null model. Taken
together these results suggest that causality in seismic dynamics
may be much broader than any normative interpretation of
``triggering''.

Our results are generally robust with respect to modifications of
the rules used to construct the network, e.g., using spatial
neighborhoods such that the construction becomes symmetric under
time reversal or taking into account magnitudes. All such
modifications have the drawback that they do not define a record
breaking process consisting of recurrences to each event. For
seismicity, our results are also unaltered if we exclude
unphysical links with propagation velocities larger than the
velocity of a ``P wave'' of about $6 km/sec$ ($\approx 0.1\%$ of
all links). This is also true if we restrict ourselves to
velocities smaller than the velocity of a shear wave of about $3.5
km/sec$ which is often thought to be more relevant.

By building certain specific features of causality into null
models, it is possible to refine predictions and examine what
features in the network of seismicity are due to those aspects of
causality and what are yet to be explained. It remains to be seen
how general our method may turn out to be. In principle it can be
applied to any high resolution data set where events occur in
space and time. Immediate applications may include analyses of
other geophysical or astrophysical data sets, brain scans
\cite{sporns04}, or analyses of models to validate or falsify them.

%
%

\begin{acknowledgments}
We thank the Southern California Earthquake Center (SCEC) for
providing the data.
\end{acknowledgments}

%
%


\begin{thebibliography}{75}
\expandafter\ifx\csname natexlab\endcsname\relax\def\natexlab#1{#1}\fi
\expandafter\ifx\csname bibnamefont\endcsname\relax
  \def\bibnamefont#1{#1}\fi
\expandafter\ifx\csname bibfnamefont\endcsname\relax
  \def\bibfnamefont#1{#1}\fi
\expandafter\ifx\csname citenamefont\endcsname\relax
  \def\citenamefont#1{#1}\fi
\expandafter\ifx\csname url\endcsname\relax
  \def\url#1{\texttt{#1}}\fi
\expandafter\ifx\csname urlprefix\endcsname\relax\def\urlprefix{URL }\fi
\providecommand{\bibinfo}[2]{#2}
\providecommand{\eprint}[2][]{\url{#2}}

\bibitem[{\citenamefont{Cote and Meisel}(1991)}]{cote91}
\bibinfo{author}{\bibfnamefont{P.~J.} \bibnamefont{Cote}} \bibnamefont{and}
  \bibinfo{author}{\bibfnamefont{L.~V.} \bibnamefont{Meisel}},
  \bibinfo{journal}{Phys. Rev. Lett.} \textbf{\bibinfo{volume}{67}},
  \bibinfo{pages}{1334} (\bibinfo{year}{1991}).

\bibitem[{\citenamefont{Sethna et~al.}(2001)\citenamefont{Sethna, Dahmen, and
  Myers}}]{sethna01}
\bibinfo{author}{\bibfnamefont{J.~P.} \bibnamefont{Sethna}},
  \bibinfo{author}{\bibfnamefont{K.~A.} \bibnamefont{Dahmen}},
  \bibnamefont{and} \bibinfo{author}{\bibfnamefont{C.~R.} \bibnamefont{Myers}},
  \bibinfo{journal}{Nature} \textbf{\bibinfo{volume}{410}},
  \bibinfo{pages}{242} (\bibinfo{year}{2001}).

\bibitem[{\citenamefont{Nagel and Paczuski}(1995)}]{nagel95}
\bibinfo{author}{\bibfnamefont{K.}~\bibnamefont{Nagel}} \bibnamefont{and}
  \bibinfo{author}{\bibfnamefont{M.}~\bibnamefont{Paczuski}},
  \bibinfo{journal}{Phys. Rev. E} \textbf{\bibinfo{volume}{51}},
  \bibinfo{pages}{2909} (\bibinfo{year}{1995}).

\bibitem[{\citenamefont{Bak}(1996)}]{bak}
\bibinfo{author}{\bibfnamefont{P.}~\bibnamefont{Bak}},
  \emph{\bibinfo{title}{How nature works}} (\bibinfo{publisher}{{Copernicus}},
  \bibinfo{address}{New York}, \bibinfo{year}{1996}).

\bibitem[{\citenamefont{Farmer et~al.}(2005)\citenamefont{Farmer, Smith, and
  Shubik}}]{farmer05}
\bibinfo{author}{\bibfnamefont{J.~D.} \bibnamefont{Farmer}},
  \bibinfo{author}{\bibfnamefont{D.~E.} \bibnamefont{Smith}}, \bibnamefont{and}
  \bibinfo{author}{\bibfnamefont{M.}~\bibnamefont{Shubik}},
  \bibinfo{journal}{Physics Today} \textbf{\bibinfo{volume}{58}},
  \bibinfo{pages}{37} (\bibinfo{year}{2005}).

\bibitem[{\citenamefont{{Bak} et~al.}(1990)\citenamefont{{Bak}, {Chen}, and
  {Tang}}}]{bak90}
\bibinfo{author}{\bibfnamefont{P.}~\bibnamefont{{Bak}}},
  \bibinfo{author}{\bibfnamefont{K.}~\bibnamefont{{Chen}}}, \bibnamefont{and}
  \bibinfo{author}{\bibfnamefont{C.}~\bibnamefont{{Tang}}},
  \bibinfo{journal}{Phys. Lett. A} \textbf{\bibinfo{volume}{147}},
  \bibinfo{pages}{297} (\bibinfo{year}{1990}).

\bibitem[{\citenamefont{Turcotte}(1999)}]{turcotte99}
\bibinfo{author}{\bibfnamefont{D.~L.} \bibnamefont{Turcotte}},
  \bibinfo{journal}{Rep. Prog. Phys.} \textbf{\bibinfo{volume}{62}},
  \bibinfo{pages}{1377} (\bibinfo{year}{1999}).

\bibitem[{\citenamefont{Bak and Sneppen}(1993)}]{bak93}
\bibinfo{author}{\bibfnamefont{P.}~\bibnamefont{Bak}} \bibnamefont{and}
  \bibinfo{author}{\bibfnamefont{K.}~\bibnamefont{Sneppen}},
  \bibinfo{journal}{Physical Review Letters} \textbf{\bibinfo{volume}{71}},
  \bibinfo{pages}{4083} (\bibinfo{year}{1993}).

\bibitem[{\citenamefont{Crutchfield and Schuster}(2003)}]{crutchfield}
\bibinfo{author}{\bibfnamefont{J.~P.} \bibnamefont{Crutchfield}}
  \bibnamefont{and} \bibinfo{author}{\bibfnamefont{P.}~\bibnamefont{Schuster}},
  \emph{\bibinfo{title}{Evolutionary Dynamics: Exploring the Interplay of
  Selection, Accident, Neutrality, and Function}} (\bibinfo{publisher}{Oxford
  University Press US}, \bibinfo{address}{NY, NY}, \bibinfo{year}{2003}).

\bibitem[{\citenamefont{Drossel}(2001)}]{drossel01}
\bibinfo{author}{\bibfnamefont{B.}~\bibnamefont{Drossel}},
  \bibinfo{journal}{Advances in Physics} \textbf{\bibinfo{volume}{50}},
  \bibinfo{pages}{209} (\bibinfo{year}{2001}).

\bibitem[{\citenamefont{Softky and Koch}(1992)}]{softky92}
\bibinfo{author}{\bibfnamefont{W.~R.} \bibnamefont{Softky}} \bibnamefont{and}
  \bibinfo{author}{\bibfnamefont{C.}~\bibnamefont{Koch}},
  \bibinfo{journal}{Neural Computation} \textbf{\bibinfo{volume}{4}},
  \bibinfo{pages}{643} (\bibinfo{year}{1992}).

\bibitem[{\citenamefont{{Hughes} et~al.}(2003)\citenamefont{{Hughes},
  {Paczuski}, {Dendy}, {Helander}, and {McClements}}}]{hughes03}
\bibinfo{author}{\bibfnamefont{D.}~\bibnamefont{{Hughes}}},
  \bibinfo{author}{\bibfnamefont{M.}~\bibnamefont{{Paczuski}}},
  \bibinfo{author}{\bibfnamefont{R.~O.} \bibnamefont{{Dendy}}},
  \bibinfo{author}{\bibfnamefont{P.}~\bibnamefont{{Helander}}},
  \bibnamefont{and} \bibinfo{author}{\bibfnamefont{K.~G.}
  \bibnamefont{{McClements}}}, \bibinfo{journal}{Phys. Rev. Lett.}
  \textbf{\bibinfo{volume}{90}}, \bibinfo{pages}{131101}
  (\bibinfo{year}{2003}).

\bibitem[{\citenamefont{{Paczuski} and {Hughes}}(2004)}]{paczuski04}
\bibinfo{author}{\bibfnamefont{M.}~\bibnamefont{{Paczuski}}} \bibnamefont{and}
  \bibinfo{author}{\bibfnamefont{D.}~\bibnamefont{{Hughes}}},
  \bibinfo{journal}{Physica A} \textbf{\bibinfo{volume}{342}},
  \bibinfo{pages}{158} (\bibinfo{year}{2004}).

\bibitem[{\citenamefont{Turcotte}(1997)}]{turcotte}
\bibinfo{author}{\bibfnamefont{D.~L.} \bibnamefont{Turcotte}},
  \emph{\bibinfo{title}{Fractals and Chaos in Geology and Geophysics}}
  (\bibinfo{publisher}{Cambridge University Press},
  \bibinfo{address}{Cambridge, UK}, \bibinfo{year}{1997}),
  \bibinfo{edition}{{2nd}} ed.

\bibitem[{\citenamefont{Stein and Wysession}(2002)}]{stein}
\bibinfo{author}{\bibfnamefont{S.}~\bibnamefont{Stein}} \bibnamefont{and}
  \bibinfo{author}{\bibfnamefont{M.}~\bibnamefont{Wysession}},
  \emph{\bibinfo{title}{An Introduction to Seismology, Earthquakes, and Earth
  Structure}} (\bibinfo{publisher}{Blackwell Publishing},
  \bibinfo{address}{Oxford, UK}, \bibinfo{year}{2002}).

\bibitem[{\citenamefont{Baiesi and Paczuski}(2005)}]{baiesi05}
\bibinfo{author}{\bibfnamefont{M.}~\bibnamefont{Baiesi}} \bibnamefont{and}
  \bibinfo{author}{\bibfnamefont{M.}~\bibnamefont{Paczuski}},
  \bibinfo{journal}{Nonlin. Proc. Geophys.} \textbf{\bibinfo{volume}{12}},
  \bibinfo{pages}{1} (\bibinfo{year}{2005}).

\bibitem[{\citenamefont{Bak et~al.}(1987)\citenamefont{Bak, Tang, and
  Wiesenfeld}}]{bak87}
\bibinfo{author}{\bibfnamefont{P.}~\bibnamefont{Bak}},
  \bibinfo{author}{\bibfnamefont{C.}~\bibnamefont{Tang}}, \bibnamefont{and}
  \bibinfo{author}{\bibfnamefont{K.}~\bibnamefont{Wiesenfeld}},
  \bibinfo{journal}{Phys. Rev. Lett.} \textbf{\bibinfo{volume}{59}},
  \bibinfo{pages}{381} (\bibinfo{year}{1987}).

\bibitem[{\citenamefont{Marzocchi et~al.}(2004)\citenamefont{Marzocchi,
  Zaccarelli, and Boschi}}]{marzocchi04}
\bibinfo{author}{\bibfnamefont{W.}~\bibnamefont{Marzocchi}},
  \bibinfo{author}{\bibfnamefont{L.}~\bibnamefont{Zaccarelli}},
  \bibnamefont{and} \bibinfo{author}{\bibfnamefont{E.}~\bibnamefont{Boschi}},
  \bibinfo{journal}{Geophys. Res. Lett.} \textbf{\bibinfo{volume}{31}},
  \bibinfo{pages}{L04601} (\bibinfo{year}{2004}).

\bibitem{note1}
In fact this is the definition of the $\epsilon$- recurrence
as discussed in Ref.~\cite{eckmann87}.

\bibitem[{\citenamefont{Helmstetter et~al.}(2003)\citenamefont{Helmstetter,
  Ouillon, and Sornette}}]{helmstetter03b}
\bibinfo{author}{\bibfnamefont{A.}~\bibnamefont{Helmstetter}},
  \bibinfo{author}{\bibfnamefont{G.}~\bibnamefont{Ouillon}}, \bibnamefont{and}
  \bibinfo{author}{\bibfnamefont{D.}~\bibnamefont{Sornette}},
  \bibinfo{journal}{J. of Geophys. Res.} \textbf{\bibinfo{volume}{108}},
  \bibinfo{pages}{2483} (\bibinfo{year}{2003}).

\bibitem[{\citenamefont{Glick}(1978)}]{glick78}
\bibinfo{author}{\bibfnamefont{N.}~\bibnamefont{Glick}}, \bibinfo{journal}{The
  American Mathematical Monthly} \textbf{\bibinfo{volume}{85}},
  \bibinfo{pages}{2} (\bibinfo{year}{1978}).

\bibitem[{\citenamefont{Nevzorov}(1987)}]{nevzorov87}
\bibinfo{author}{\bibfnamefont{V.~B.} \bibnamefont{Nevzorov}},
  \bibinfo{journal}{Theory Prob. Appl.} \textbf{\bibinfo{volume}{32}},
  \bibinfo{pages}{201} (\bibinfo{year}{1987}).

\bibitem[{\citenamefont{Nevzorov}(2001)}]{nevzorov}
\bibinfo{author}{\bibfnamefont{V.~B.} \bibnamefont{Nevzorov}},
  \emph{\bibinfo{title}{Records: Mathematical theory}}, vol.
  \bibinfo{volume}{194} of \emph{\bibinfo{series}{Translations of Mathematical
  Monographs}} (\bibinfo{publisher}{American Mathematical Society},
  \bibinfo{address}{Providence, Rhode Island}, \bibinfo{year}{2001}).

\bibitem{catalog}
The seismic catalog was obtained from
http://www.data.scec.org/ftp/catalogs/SHLK/.

\bibitem[{\citenamefont{Baiesi and Paczuski}(2004)}]{baiesi04}
\bibinfo{author}{\bibfnamefont{M.}~\bibnamefont{Baiesi}} \bibnamefont{and}
  \bibinfo{author}{\bibfnamefont{M.}~\bibnamefont{Paczuski}},
  \bibinfo{journal}{Phys. Rev. E} \textbf{\bibinfo{volume}{69}},
  \bibinfo{pages}{066106} (\bibinfo{year}{2004}).

\bibitem[{\citenamefont{Baiesi}(2006)}]{baiesi06}
\bibinfo{author}{\bibfnamefont{M.}~\bibnamefont{Baiesi}},
  \bibinfo{journal}{Physica A} \textbf{\bibinfo{volume}{360}},
  \bibinfo{pages}{534} (\bibinfo{year}{2006}).

\bibitem[{\citenamefont{Albert and Barabasi}(2002)}]{albert02}
\bibinfo{author}{\bibfnamefont{R.}~\bibnamefont{Albert}} \bibnamefont{and}
  \bibinfo{author}{\bibfnamefont{A.-L.} \bibnamefont{Barabasi}},
  \bibinfo{journal}{Rev. Mod. Phys.} \textbf{\bibinfo{volume}{74}},
  \bibinfo{pages}{47} (\bibinfo{year}{2002}).

\bibitem[{\citenamefont{Newman}(2003)}]{newman03}
\bibinfo{author}{\bibfnamefont{M.~E.~J.} \bibnamefont{Newman}},
  \bibinfo{journal}{SIAM Review} \textbf{\bibinfo{volume}{45}},
  \bibinfo{pages}{167} (\bibinfo{year}{2003}).

\bibitem[{\citenamefont{Davidsen et~al.}(2006)\citenamefont{Davidsen,
  Grassberger, and Paczuski}}]{davidsen05pm}
\bibinfo{author}{\bibfnamefont{J.}~\bibnamefont{Davidsen}},
  \bibinfo{author}{\bibfnamefont{P.}~\bibnamefont{Grassberger}},
  \bibnamefont{and} \bibinfo{author}{\bibfnamefont{M.}~\bibnamefont{Paczuski}},
  \bibinfo{journal}{Geophys. Res. Lett.} \textbf{\bibinfo{volume}{33}},
  \bibinfo{pages}{L11304} (\bibinfo{year}{2006}).

\bibitem[{\citenamefont{Shreim et~al.}(2006)\citenamefont{Shreim, Grassberger,
  Nadler, Samuelsson, Socolar, and Paczuski}}]{shreim06}
\bibinfo{author}{\bibfnamefont{A.}~\bibnamefont{Shreim}},
  \bibinfo{author}{\bibfnamefont{P.}~\bibnamefont{Grassberger}},
  \bibinfo{author}{\bibfnamefont{W.}~\bibnamefont{Nadler}},
  \bibinfo{author}{\bibfnamefont{B.}~\bibnamefont{Samuelsson}},
  \bibinfo{author}{\bibfnamefont{J.~E.~S.} \bibnamefont{Socolar}},
  \bibnamefont{and} \bibinfo{author}{\bibfnamefont{M.}~\bibnamefont{Paczuski}}
  (\bibinfo{year}{2006}), \bibinfo{note}{to be published in Phys. Rev. Lett.}.

\bibitem[{\citenamefont{Mandelbrot}(1995)}]{mandelbrot95}
\bibinfo{author}{\bibfnamefont{B.~B.} \bibnamefont{Mandelbrot}}, in
  \emph{\bibinfo{booktitle}{Fractal Geometry and Stochastics}}, edited by
  \bibinfo{editor}{\bibfnamefont{C.}~\bibnamefont{Bandt}},
  \bibinfo{editor}{\bibfnamefont{S.}~\bibnamefont{Graf}}, \bibnamefont{and}
  \bibinfo{editor}{\bibfnamefont{M.}~\bibnamefont{Z{\"a}hle}}
  (\bibinfo{publisher}{Birkh{\"a}user Verlag}, \bibinfo{address}{Basel},
  \bibinfo{year}{1995}).

\bibitem[{\citenamefont{Sibani et~al.}(1998)\citenamefont{Sibani, Brandt, and
  Alstr{\o}m}}]{sibani98}
\bibinfo{author}{\bibfnamefont{P.}~\bibnamefont{Sibani}},
  \bibinfo{author}{\bibfnamefont{M.}~\bibnamefont{Brandt}}, \bibnamefont{and}
  \bibinfo{author}{\bibfnamefont{P.}~\bibnamefont{Alstr{\o}m}}, \bibinfo{journal}{Int. J. Mod.
  Phys.} \textbf{\bibinfo{volume}{12}}, \bibinfo{pages}{361}
  (\bibinfo{year}{1998}).

\bibitem[{\citenamefont{Krug and Jain}(2005)}]{krug05}
\bibinfo{author}{\bibfnamefont{J.}~\bibnamefont{Krug}} \bibnamefont{and}
  \bibinfo{author}{\bibfnamefont{K.}~\bibnamefont{Jain}},
  \bibinfo{journal}{Physica A} \textbf{\bibinfo{volume}{358}},
  \bibinfo{pages}{1} (\bibinfo{year}{2005}).

\bibitem[{\citenamefont{Rundle et~al.}(2003)\citenamefont{Rundle, Turcotte,
  Shcherbakov, Klein, and Sammis}}]{rundle03}
\bibinfo{author}{\bibfnamefont{J.~B.} \bibnamefont{Rundle}},
  \bibinfo{author}{\bibfnamefont{D.~L.} \bibnamefont{Turcotte}},
  \bibinfo{author}{\bibfnamefont{R.}~\bibnamefont{Shcherbakov}},
  \bibinfo{author}{\bibfnamefont{W.}~\bibnamefont{Klein}}, \bibnamefont{and}
  \bibinfo{author}{\bibfnamefont{C.}~\bibnamefont{Sammis}},
  \bibinfo{journal}{Review of Geophysics} \textbf{\bibinfo{volume}{41}},
  \bibinfo{pages}{1019} (\bibinfo{year}{2003}).

\bibitem[{\citenamefont{Davidsen and Paczuski}(2005)}]{davidsen05m}
\bibinfo{author}{\bibfnamefont{J.}~\bibnamefont{Davidsen}} \bibnamefont{and}
  \bibinfo{author}{\bibfnamefont{M.}~\bibnamefont{Paczuski}},
  \bibinfo{journal}{Phys. Rev. Lett.} \textbf{\bibinfo{volume}{94}},
  \bibinfo{pages}{048501} (\bibinfo{year}{2005}).

\bibitem[{\citenamefont{Davidsen and Goltz}(2004)}]{davidsen04}
\bibinfo{author}{\bibfnamefont{J.}~\bibnamefont{Davidsen}} \bibnamefont{and}
  \bibinfo{author}{\bibfnamefont{C.}~\bibnamefont{Goltz}},
  \bibinfo{journal}{Geophys. Res. Lett.} \textbf{\bibinfo{volume}{31}},
  \bibinfo{pages}{L21612} (\bibinfo{year}{2004}).

\bibitem[{\citenamefont{Bak et~al.}(2002)\citenamefont{Bak, Christensen, Danon,
  and Scanlon}}]{bak02}
\bibinfo{author}{\bibfnamefont{P.}~\bibnamefont{Bak}},
  \bibinfo{author}{\bibfnamefont{K.}~\bibnamefont{Christensen}},
  \bibinfo{author}{\bibfnamefont{L.}~\bibnamefont{Danon}}, \bibnamefont{and}
  \bibinfo{author}{\bibfnamefont{T.}~\bibnamefont{Scanlon}},
  \bibinfo{journal}{Phys. Rev. Lett.} \textbf{\bibinfo{volume}{88}},
  \bibinfo{pages}{178501} (\bibinfo{year}{2002}).

\bibitem[{\citenamefont{Corral}(2003)}]{corral03}
\bibinfo{author}{\bibfnamefont{A.}~\bibnamefont{Corral}},
  \bibinfo{journal}{Phys. Rev. E} \textbf{\bibinfo{volume}{68}},
  \bibinfo{pages}{035102} (\bibinfo{year}{2003}).

\bibitem[{\citenamefont{Corral}(2004)}]{corral04}
\bibinfo{author}{\bibfnamefont{A.}~\bibnamefont{Corral}},
  \bibinfo{journal}{Phys. Rev. Lett.} \textbf{\bibinfo{volume}{92}},
  \bibinfo{pages}{108501} (\bibinfo{year}{2004}).

\bibitem[{\citenamefont{Shearer et~al.}(2003)\citenamefont{Shearer, Hauksson,
  Lin, and Kilb}}]{shearer03}
\bibinfo{author}{\bibfnamefont{P.}~\bibnamefont{Shearer}},
  \bibinfo{author}{\bibfnamefont{E.}~\bibnamefont{Hauksson}},
  \bibinfo{author}{\bibfnamefont{G.}~\bibnamefont{Lin}}, \bibnamefont{and}
  \bibinfo{author}{\bibfnamefont{D.}~\bibnamefont{Kilb}}, \bibinfo{journal}{Eos
  Trans. AGU} \textbf{\bibinfo{volume}{84}}, \bibinfo{pages}{46}
  (\bibinfo{year}{2003}).

\bibitem[{\citenamefont{Shearer et~al.}(2005)\citenamefont{Shearer, Hauksson,
  and Lin}}]{shearer05}
\bibinfo{author}{\bibfnamefont{P.}~\bibnamefont{Shearer}},
  \bibinfo{author}{\bibfnamefont{E.}~\bibnamefont{Hauksson}}, \bibnamefont{and}
  \bibinfo{author}{\bibfnamefont{G.}~\bibnamefont{Lin}},
  \bibinfo{journal}{Bull. Seismol. Soc. America} \textbf{\bibinfo{volume}{95}},
  \bibinfo{pages}{904} (\bibinfo{year}{2005}).

\bibitem[{\citenamefont{Wiemer and Wyss}(2000)}]{wiemer00}
\bibinfo{author}{\bibfnamefont{S.}~\bibnamefont{Wiemer}} \bibnamefont{and}
  \bibinfo{author}{\bibfnamefont{M.}~\bibnamefont{Wyss}},
  \bibinfo{journal}{Bull. Seismol. Soc. America} \textbf{\bibinfo{volume}{90}},
  \bibinfo{pages}{859} (\bibinfo{year}{2000}).

\bibitem[{\citenamefont{Schwartz and Coppersmith}(1984)}]{schwartz84}
\bibinfo{author}{\bibfnamefont{D.~P.}~\bibnamefont{Schwartz}} \bibnamefont{and}
  \bibinfo{author}{\bibfnamefont{K.~J.}~\bibnamefont{Coppersmith}},
  \bibinfo{journal}{J. of Geophys. Res.} \textbf{\bibinfo{volume}{89}},
  \bibinfo{pages}{5681} (\bibinfo{year}{1984}).

\bibitem[{\citenamefont{Stirling et~al.}(1996)\citenamefont{Stirling, Wesnousky, and Shimazaki}}]{stirling96}
\bibinfo{author}{\bibfnamefont{M.~W.}~\bibnamefont{Stirling}},
  \bibinfo{author}{\bibfnamefont{S.~G.}~\bibnamefont{Wesnousky}}, \bibnamefont{and}
  \bibinfo{author}{\bibfnamefont{K.}~\bibnamefont{Shimazaki}},
  \bibinfo{journal}{Geophys. J. Int.} \textbf{\bibinfo{volume}{124}},
  \bibinfo{pages}{833} (\bibinfo{year}{1996}).

\bibitem[{\citenamefont{Matthews et~al.}(2002)\citenamefont{Matthews, Ellsworth, and Reasenberg}}]{matthews02}
\bibinfo{author}{\bibfnamefont{M.~V.}~\bibnamefont{Matthews}},
  \bibinfo{author}{\bibfnamefont{W.~L.}~\bibnamefont{Ellsworth}}, \bibnamefont{and}
  \bibinfo{author}{\bibfnamefont{P.~A.}~\bibnamefont{Reasenberg}},
  \bibinfo{journal}{Bull. Seismol. Soc. America} \textbf{\bibinfo{volume}{92}},
  \bibinfo{pages}{2233} (\bibinfo{year}{2002}).

\bibitem[{\citenamefont{Omori}(1894)}]{omori}
\bibinfo{author}{\bibfnamefont{F.}~\bibnamefont{Omori}},
  \bibinfo{journal}{Journal of College Science, Imperial University of Tokyo}
  \textbf{\bibinfo{volume}{7}}, \bibinfo{pages}{111} (\bibinfo{year}{1894}).

\bibitem[{\citenamefont{Rubin and Gillard}(2000)}]{rubin00}
\bibinfo{author}{\bibfnamefont{A.~M.} \bibnamefont{Rubin}} \bibnamefont{and}
  \bibinfo{author}{\bibfnamefont{D.}~\bibnamefont{Gillard}},
  \bibinfo{journal}{Journal of Geophysical Research}
  \textbf{\bibinfo{volume}{105}}, \bibinfo{pages}{19095}
  (\bibinfo{year}{2000}).

\bibitem[{\citenamefont{Kagan}(2002)}]{kagan02}
\bibinfo{author}{\bibfnamefont{Y.~Y.} \bibnamefont{Kagan}},
  \bibinfo{journal}{Bull. Seismol. Soc. America} \textbf{\bibinfo{volume}{92}},
  \bibinfo{pages}{641} (\bibinfo{year}{2002}).

\bibitem[{\citenamefont{Wells and Coppersmith}(1994)}]{wells94}
\bibinfo{author}{\bibfnamefont{D.~L.} \bibnamefont{Wells}} \bibnamefont{and}
  \bibinfo{author}{\bibfnamefont{K.~J.} \bibnamefont{Coppersmith}},
  \bibinfo{journal}{Bull. Seismol. Soc. America} \textbf{\bibinfo{volume}{84}},
  \bibinfo{pages}{974} (\bibinfo{year}{1994}).

\bibitem{note2}
Note that a systematic
dependence of the location error on magnitude has not been reported
in the literature and is also not present in the catalog at hand.
It is unlikely that the characteristic length we see ($l^*(m)$) is merely
an artifact due to location error growing with magnitude.

\bibitem{note3}
For $m=2.5$, the hypothesis that the out-degree
distribution is Poissonian is
rejected by the $\chi^2$-test at the $99.5\%$
significance level. Specifically, we find $\chi_0^2 =
66.7 \times 10^3$ for 22 degrees of freedom.

\bibitem{note4}
For $m=2.5$, the hypothesis that the in-degree
distribution is Poissonian is
rejected by the $\chi^2$-test at the $99.5\%$
significance level. Specifically, we find $\chi_0^2 = 88.7$
for 15 degrees of freedom.

\bibitem{note5}
The averages were performed over all \emph{links}
emanating from events with fixed in-degree or out-degree,
respectively.

\bibitem[{\citenamefont{Soffer and V{\'a}zquez}(2005)}]{soffer05}
\bibinfo{author}{\bibfnamefont{S.~N.} \bibnamefont{Soffer}} \bibnamefont{and}
  \bibinfo{author}{\bibfnamefont{A.}~\bibnamefont{V{\'a}zquez}},
  \bibinfo{journal}{Phys. Rev. E} \textbf{\bibinfo{volume}{71}},
  \bibinfo{pages}{057101} (\bibinfo{year}{2005}).

\bibitem[{\citenamefont{Helmstetter et~al.}(2005)\citenamefont{Helmstetter,
  Kagan, and Jackson}}]{helmstetter05}
\bibinfo{author}{\bibfnamefont{A.}~\bibnamefont{Helmstetter}},
  \bibinfo{author}{\bibfnamefont{Y.~Y.} \bibnamefont{Kagan}}, \bibnamefont{and}
  \bibinfo{author}{\bibfnamefont{D.~D.} \bibnamefont{Jackson}},
  \bibinfo{journal}{J. Geophys. Res.}
  \textbf{\bibinfo{volume}{110}} (\bibinfo{year}{2005}).

\bibitem[{\citenamefont{Harris}(1998)}]{harris98}
\bibinfo{author}{\bibfnamefont{R.~A.} \bibnamefont{Harris}},
  \bibinfo{journal}{J. Geophys. Res.}
  \textbf{\bibinfo{volume}{103}}, \bibinfo{pages}{24347}
  (\bibinfo{year}{1998}).

\bibitem[{\citenamefont{Stein}(1999)}]{stein99}
\bibinfo{author}{\bibfnamefont{R.~S.} \bibnamefont{Stein}},
  \bibinfo{journal}{Nature (London)} \textbf{\bibinfo{volume}{402}},
  \bibinfo{pages}{605} (\bibinfo{year}{1999}).

\bibitem[{\citenamefont{Freed}(2005)}]{freed05}
\bibinfo{author}{\bibfnamefont{A.~M.} \bibnamefont{Freed}},
  \bibinfo{journal}{Ann. Rev. Earth Planet. Sci.}
  \textbf{\bibinfo{volume}{33}}, \bibinfo{pages}{335} (\bibinfo{year}{2005}).

\bibitem[{\citenamefont{Main}(2006)}]{main06}
\bibinfo{author}{\bibfnamefont{I.}~\bibnamefont{Main}},
  \bibinfo{journal}{Nature (London)} \textbf{\bibinfo{volume}{441}},
  \bibinfo{pages}{704} (\bibinfo{year}{2006}).

\bibitem[{\citenamefont{Mallman and Zoback}(2007)}]{mallman07}
\bibinfo{author}{\bibfnamefont{E.~P.} \bibnamefont{Mallman}} \bibnamefont{and}
  \bibinfo{author}{\bibfnamefont{M.~D.} \bibnamefont{Zoback}},
  \bibinfo{journal}{J. Geophys. Res.}
  \textbf{\bibinfo{volume}{112}}, \bibinfo{pages}{B03304}
  (\bibinfo{year}{2007}).

\bibitem[{\citenamefont{Dietrich}(1994)}]{dietrich94}
\bibinfo{author}{\bibfnamefont{J.}~\bibnamefont{Dietrich}},
  \bibinfo{journal}{J. Geophys. Res.}
  \textbf{\bibinfo{volume}{99}}, \bibinfo{pages}{2601} (\bibinfo{year}{1994}).

\bibitem[{\citenamefont{Das and Scholz}(1981)}]{das81}
\bibinfo{author}{\bibfnamefont{S.}~\bibnamefont{Das}} \bibnamefont{and}
  \bibinfo{author}{\bibfnamefont{C.~H.} \bibnamefont{Scholz}},
  \bibinfo{journal}{J. Geophys. Res.}
  \textbf{\bibinfo{volume}{86}}, \bibinfo{pages}{6039} (\bibinfo{year}{1981}).

\bibitem[{\citenamefont{Shaw}(1993)}]{shaw93}
\bibinfo{author}{\bibfnamefont{B.~E.} \bibnamefont{Shaw}},
  \bibinfo{journal}{Geophys. Res. Lett.}
  \textbf{\bibinfo{volume}{20}}, \bibinfo{pages}{907} (\bibinfo{year}{1993}).

\bibitem[{\citenamefont{Main}(2000)}]{main00}
\bibinfo{author}{\bibfnamefont{I.}~\bibnamefont{Main}},
  \bibinfo{journal}{Bull. Seismol. Soc. America}
  \textbf{\bibinfo{volume}{90}}, \bibinfo{pages}{86} (\bibinfo{year}{2000}).

\bibitem[{\citenamefont{Mikumo and Miyatake}(1979)}]{mikumo79}
\bibinfo{author}{\bibfnamefont{T.}~\bibnamefont{Mikumo}} \bibnamefont{and}
  \bibinfo{author}{\bibfnamefont{T.}~\bibnamefont{Miyatake}},
  \bibinfo{journal}{Geophys. J. Royal Astr. Soc.}
  \textbf{\bibinfo{volume}{59}}, \bibinfo{pages}{497} (\bibinfo{year}{1979}).

\bibitem[{\citenamefont{Scholz}(1968)}]{scholz68}
\bibinfo{author}{\bibfnamefont{C.~H.} \bibnamefont{Scholz}},
  \bibinfo{journal}{J. Geophys. Res.}
  \textbf{\bibinfo{volume}{73}}, \bibinfo{pages}{1417} (\bibinfo{year}{1968}).

\bibitem[{\citenamefont{Nur and Booker}(1972)}]{nur72}
\bibinfo{author}{\bibfnamefont{A.}~\bibnamefont{Nur}} \bibnamefont{and}
  \bibinfo{author}{\bibfnamefont{J.~R.} \bibnamefont{Booker}},
  \bibinfo{journal}{Science}
  \textbf{\bibinfo{volume}{175}}, \bibinfo{pages}{885} (\bibinfo{year}{1972}).

\bibitem[{\citenamefont{Hergarten and Neugebauer}(2002)}]{hergarten02}
\bibinfo{author}{\bibfnamefont{S.}~\bibnamefont{Hergarten}} \bibnamefont{and}
  \bibinfo{author}{\bibfnamefont{H.~J.} \bibnamefont{Neugebauer}},
  \bibinfo{journal}{Phys. Rev. Lett.} \textbf{\bibinfo{volume}{88}},
  \bibinfo{pages}{238501} (\bibinfo{year}{2002}).

\bibitem[{\citenamefont{King and Cocco}(2001)}]{king01}
\bibinfo{author}{\bibfnamefont{G.~C.~P.} \bibnamefont{King}} \bibnamefont{and}
  \bibinfo{author}{\bibfnamefont{M.}~\bibnamefont{Cocco}},
  \bibinfo{journal}{Adv. Geophys.} \textbf{\bibinfo{volume}{44}},
  \bibinfo{pages}{1} (\bibinfo{year}{2001}).

\bibitem[{\citenamefont{Parsons}(2002)}]{parsons02}
\bibinfo{author}{\bibfnamefont{T.}~\bibnamefont{Parsons}},
  \bibinfo{journal}{J. Geophys. Res.}
  \textbf{\bibinfo{volume}{107}}, \bibinfo{pages}{2199} (\bibinfo{year}{2002}).

\bibitem[{\citenamefont{Marsan}(2003)}]{marsan03a}
\bibinfo{author}{\bibfnamefont{D.}~\bibnamefont{Marsan}},
  \bibinfo{journal}{J. Geophys. Res.}
  \textbf{\bibinfo{volume}{108}}, \bibinfo{pages}{2266} (\bibinfo{year}{2003}).

\bibitem[{\citenamefont{Felzer and Brodsky}(2005)}]{felzer03}
\bibinfo{author}{\bibfnamefont{K.~R.} \bibnamefont{Felzer}} \bibnamefont{and}
  \bibinfo{author}{\bibfnamefont{E.~E.} \bibnamefont{Brodsky}},
  \bibinfo{journal}{J. Geophys. Res.}
  \textbf{\bibinfo{volume}{110}} (\bibinfo{year}{2005}).

\bibitem[{\citenamefont{Steacy et~al.}(2004)\citenamefont{Steacy, Marsan,
  Nalbant, and McCloskey}}]{steacy04}
\bibinfo{author}{\bibfnamefont{S.}~\bibnamefont{Steacy}},
  \bibinfo{author}{\bibfnamefont{D.}~\bibnamefont{Marsan}},
  \bibinfo{author}{\bibfnamefont{S.~S.} \bibnamefont{Nalbant}},
  \bibnamefont{and}
  \bibinfo{author}{\bibfnamefont{J.}~\bibnamefont{McCloskey}},
  \bibinfo{journal}{J. Geophys. Res.}
  \textbf{\bibinfo{volume}{109}} (\bibinfo{year}{2004}).

\bibitem[{\citenamefont{Felzer et~al.}(2002)\citenamefont{Felzer, Becker,
  Abercrombie, Ekstr{\"o}m, and Rice}}]{felzer02}
\bibinfo{author}{\bibfnamefont{K.~R.} \bibnamefont{Felzer}},
  \bibinfo{author}{\bibfnamefont{T.~W.} \bibnamefont{Becker}},
  \bibinfo{author}{\bibfnamefont{R.~E.} \bibnamefont{Abercrombie}},
  \bibinfo{author}{\bibfnamefont{G.}~\bibnamefont{Ekstr{\"o}m}},
  \bibnamefont{and} \bibinfo{author}{\bibfnamefont{J.~R.} \bibnamefont{Rice}},
  \bibinfo{journal}{J. Geophys. Res.}
  \textbf{\bibinfo{volume}{107}}, \bibinfo{pages}{2190} (\bibinfo{year}{2002}).

\bibitem[{\citenamefont{Kilb et~al.}(2000)\citenamefont{Kilb, Gomberg, and
  Bodin}}]{kilb00}
\bibinfo{author}{\bibfnamefont{D.}~\bibnamefont{Kilb}},
  \bibinfo{author}{\bibfnamefont{J.~S.} \bibnamefont{Gomberg}},
  \bibnamefont{and} \bibinfo{author}{\bibfnamefont{P.}~\bibnamefont{Bodin}},
  \bibinfo{journal}{Nature (London)} \textbf{\bibinfo{volume}{408}},
  \bibinfo{pages}{570} (\bibinfo{year}{2000}).

\bibitem[{\citenamefont{Gomberg et~al.}(2003)\citenamefont{Gomberg, Bodin, and
  Reasenberg}}]{gomberg03}
\bibinfo{author}{\bibfnamefont{J.~S.} \bibnamefont{Gomberg}},
  \bibinfo{author}{\bibfnamefont{P.}~\bibnamefont{Bodin}}, \bibnamefont{and}
  \bibinfo{author}{\bibfnamefont{P.~A.} \bibnamefont{Reasenberg}},
  \bibinfo{journal}{Bull. Seismol. Soc. America}
  \textbf{\bibinfo{volume}{93}}, \bibinfo{pages}{118} (\bibinfo{year}{2003}).

\bibitem[{\citenamefont{Kilb}(2003)}]{kilb03}
\bibinfo{author}{\bibfnamefont{D.}~\bibnamefont{Kilb}},
  \bibinfo{journal}{J. Geophys. Res.}
  \textbf{\bibinfo{volume}{108}}, \bibinfo{pages}{2012} (\bibinfo{year}{2003}).

\bibitem[{\citenamefont{Parsons}(2005)}]{parsons05}
\bibinfo{author}{\bibfnamefont{T.}~\bibnamefont{Parsons}},
  \bibinfo{journal}{Geophys. Res. Lett.}
  \textbf{\bibinfo{volume}{32}}, \bibinfo{pages}{L04302}
  (\bibinfo{year}{2005}).

\bibitem[{\citenamefont{Johnson and Jia}(2005)}]{johnson05}
\bibinfo{author}{\bibfnamefont{P.~A.} \bibnamefont{Johnson}} \bibnamefont{and}
  \bibinfo{author}{\bibfnamefont{X.}~\bibnamefont{Jia}},
  \bibinfo{journal}{Nature (London)} \textbf{\bibinfo{volume}{437}},
  \bibinfo{pages}{871} (\bibinfo{year}{2005}).

\bibitem[{\citenamefont{Pollitz and Johnston}(2006)}]{pollitz06}
\bibinfo{author}{\bibfnamefont{F.~F.} \bibnamefont{Pollitz}} \bibnamefont{and}
  \bibinfo{author}{\bibfnamefont{M.~J.~S.} \bibnamefont{Johnston}},
  \bibinfo{journal}{Geophys. Res. Lett.}
  \textbf{\bibinfo{volume}{33}}, \bibinfo{pages}{L15318}
  (\bibinfo{year}{2006}).

\bibitem[{\citenamefont{Felzer and Brodsky}(2006)}]{felzer06}
\bibinfo{author}{\bibfnamefont{K.~R.} \bibnamefont{Felzer}} \bibnamefont{and}
  \bibinfo{author}{\bibfnamefont{E.~E.} \bibnamefont{Brodsky}},
  \bibinfo{journal}{Nature (London)} \textbf{\bibinfo{volume}{441}},
  \bibinfo{pages}{735} (\bibinfo{year}{2006}).

\bibitem[{\citenamefont{Kagan and Knopoff}(1987)}]{kagan87}
\bibinfo{author}{\bibfnamefont{Y.~Y.} \bibnamefont{Kagan}} \bibnamefont{and}
  \bibinfo{author}{\bibfnamefont{L.}~\bibnamefont{Knopoff}},
  \bibinfo{journal}{Science}
  \textbf{\bibinfo{volume}{236}}, \bibinfo{pages}{1563} (\bibinfo{year}{1987}).

\bibitem[{\citenamefont{Ogata}(1988)}]{ogata88}
\bibinfo{author}{\bibfnamefont{Y.}~\bibnamefont{Ogata}},
  \bibinfo{journal}{J. American Stat. Assoc.}
  \textbf{\bibinfo{volume}{83}}, \bibinfo{pages}{9} (\bibinfo{year}{1988}).

\bibitem[{\citenamefont{Helmstetter et~al.}(2006)\citenamefont{Helmstetter,
  Kagan, and Jackson}}]{helmstetter06}
\bibinfo{author}{\bibfnamefont{A.}~\bibnamefont{Helmstetter}},
  \bibinfo{author}{\bibfnamefont{Y.~Y.} \bibnamefont{Kagan}}, \bibnamefont{and}
  \bibinfo{author}{\bibfnamefont{D.~D.} \bibnamefont{Jackson}},
  \bibinfo{journal}{Bull. Seismol. Soc. America}
  \textbf{\bibinfo{volume}{96}}, \bibinfo{pages}{90} (\bibinfo{year}{2006}).

\bibitem[{\citenamefont{Shcherbakov et~al.}(2005)\citenamefont{Shcherbakov,
  Yakovlev, Turcotte, and Rundle}}]{shcherbakov05}
\bibinfo{author}{\bibfnamefont{R.}~\bibnamefont{Shcherbakov}},
  \bibinfo{author}{\bibfnamefont{G.}~\bibnamefont{Yakovlev}},
  \bibinfo{author}{\bibfnamefont{D.~L.} \bibnamefont{Turcotte}},
  \bibnamefont{and} \bibinfo{author}{\bibfnamefont{J.~B.}
  \bibnamefont{Rundle}}, \bibinfo{journal}{Phys. Rev. Lett.}
  \textbf{\bibinfo{volume}{95}}, \bibinfo{pages}{218501}
  (\bibinfo{year}{2005}).

\bibitem[{\citenamefont{Shcherbakov et~al.}(2006)\citenamefont{Shcherbakov,
  Turcotte, and Rundle}}]{shcherbakov06}
\bibinfo{author}{\bibfnamefont{R.}~\bibnamefont{Shcherbakov}},
  \bibinfo{author}{\bibfnamefont{D.~L.} \bibnamefont{Turcotte}},
  \bibnamefont{and} \bibinfo{author}{\bibfnamefont{J.~B.}
  \bibnamefont{Rundle}}, \bibinfo{journal}{Bull. Seismol.
  Soc. America} \textbf{\bibinfo{volume}{96}}, \bibinfo{pages}{376}
  (\bibinfo{year}{2006}).

\bibitem[{\citenamefont{Sporns et~al.}(2004)\citenamefont{Sporns, Chialvo,
  Kaiser, and Hilgetag}}]{sporns04}
\bibinfo{author}{\bibfnamefont{O.}~\bibnamefont{Sporns}},
  \bibinfo{author}{\bibfnamefont{D.~R.} \bibnamefont{Chialvo}},
  \bibinfo{author}{\bibfnamefont{M.}~\bibnamefont{Kaiser}}, \bibnamefont{and}
  \bibinfo{author}{\bibfnamefont{C.~C.} \bibnamefont{Hilgetag}},
  \bibinfo{journal}{Trends in Cognitive Sciences} \textbf{\bibinfo{volume}{8}},
  \bibinfo{pages}{418} (\bibinfo{year}{2004}).

\bibitem[{\citenamefont{Eckmann et~al.}(1987)\citenamefont{Eckmann, Kamphorst,
  and Ruelle}}]{eckmann87}
\bibinfo{author}{\bibfnamefont{J.-P.} \bibnamefont{Eckmann}},
  \bibinfo{author}{\bibfnamefont{S.~O.} \bibnamefont{Kamphorst}},
  \bibnamefont{and} \bibinfo{author}{\bibfnamefont{D.}~\bibnamefont{Ruelle}},
  \bibinfo{journal}{Europhys. Lett.} \textbf{\bibinfo{volume}{4}},
  \bibinfo{pages}{973} (\bibinfo{year}{1987}).

\end{thebibliography}


\end{document}